\documentclass[%
      aps,
      prd,
      floatfix,
      preprintnumbers,
      preprint,
      showpacs,
      nofootinbib,
      tightenlines,
      amssymb,
      amsmath
]{revtex4-1}

\usepackage{ifpdf}
\ifpdf
	\usepackage{cmap} 
	\usepackage[pdftex]{color,graphicx}
	\usepackage{epstopdf} 
\else
	\usepackage{color,graphicx}
\fi
\usepackage{bm}
\usepackage[dvipsnames]{xcolor}
\usepackage{hyperref}
\hypersetup{%
colorlinks=true,%
citecolor=Blue,%
filecolor=Black,%
linkcolor=Blue,%
urlcolor=Violet
}
\ifpdf
\usepackage[activate={true,nocompatibility},final,tracking=true,kerning=true,spacing=true,factor=1100,stretch=10,shrink=10]{microtype}
\fi

\newcommand{\ud}     {\mathrm{d}}
\newcommand{\gev}    {\:\mathrm{GeV}}
\newcommand{\mev}    {\:\mathrm{MeV}}
\newcommand{\gevsq}  {\:\mathrm{GeV}^2}

\newcommand{\ceps}{\varepsilon}

\newcommand{\average}[1]{\left\langle{#1}\right\rangle}

\newcommand{\eq}[1]{Eq.(\ref{#1})}

\begin{document}
\title{%
Nuclear Effects in the Deuteron and Constraints on the $d/u$ Ratio  
}

\author{S.~I.~Alekhin}
\email[]{sergey.alekhin@ihep.ru}
\affiliation{II. Institut f\"ur Theoretische Physik, Universit\"at Hamburg, \\
  Luruper Chaussee 149, D--22761 Hamburg, Germany}
\affiliation{Institute of High Energy Physics, 142281 Protvino, Moscow Region, Russia}
\author{S.~A.~Kulagin}
\email[]{kulagin@ms2.inr.ac.ru}
\affiliation{Institute for Nuclear Research of the Russian Academy of Sciences,
60-letiya Oktyabrya prospekt 7a, 117312 Moscow, Russia} 
\author{R.~Petti}
\email[]{Roberto.Petti@cern.ch}
\affiliation{Department of Physics and Astronomy, University of South Carolina, Columbia, South Carolina 29208, USA}


\begin{abstract}
We present a detailed study of nuclear corrections in the deuteron (D) by performing an analysis of 
data from deep inelastic scattering off proton and D, dilepton pair production in $pp$ and $p{\rm D}$ 
interactions, and $W^\pm$ and $Z$ boson production in $pp$ and $p \bar p$ collisions. 
In particular, we discuss the determination of the off-shell function describing the modification 
of the parton distribution functions in bound nucleons in the context of global QCD fits. 
Our results are consistent with the ones obtained independently from the study of data on deep 
inelastic scattering off heavy nuclei with mass number $A\geq4$, further confirming the universality 
of the off-shell function of the bound nucleon. 
We also study the sensitivity to the modeling of the deuteron wave function. 
As an important application we discuss the impact of nuclear corrections to the deuteron 
on the determination of the $d$ quark distribution.

\end{abstract}


\pacs{13.60.Hb, 12.38.Qk} 
\maketitle

\section{Introduction}
\label{sec:intro}

Parton distribution functions (PDFs) are universal process-independent characteristics
of the target, which determine leading contributions to the cross sections of various hard 
processes involving leptons and hadrons~\cite{Collins:1989gx}. The PDF content of  
both the proton and the neutron is extracted from global fits 
\cite{Alekhin:2017kpj,Harland-Lang:2014zoa,Ball:2014uwa,Dulat:2015mca} 
to experimental data at large momentum transfer, including lepton deep inelastic scattering (DIS), 
lepton-pair production (Drell-Yan process), jet production, and  
$W$ and $Z$ boson production in hadron collisions.  
In order to disentangle the content of different parton flavors, global fits must include 
complementary data which are flavor dependent. 
Traditionally, the most efficient separation between $d$ and $u$ quark
distributions is obtained by comparing charged-lepton DIS data for proton and deuterium targets, 
the latter being considered as an ``effective" neutron target. 
Since the deuteron is a weakly bound nucleus with a binding energy of about $2.2\mev$ --
accounting only for about 0.1\% of its mass -- it is often assumed to be well approximated by the sum 
of a quasi-free proton and a quasi-free neutron in the PDF analyses.

However, charged-lepton DIS data from various nuclear targets demonstrate significant nuclear effects
with a rate that is more than 1 order of magnitude larger than the ratio of the nuclear
binding energy to the nucleon mass (for a review see Refs.~\cite{Arneodo:1992wf,Norton:2003cb}).
These observations indicate that the nuclear environment plays an important role 
even at energies and momenta much higher
than those involved in typical nuclear ground state processes
\cite{Arneodo:1992wf,Norton:2003cb,Geesaman:1995yd}.
Similar considerations can be drawn for the DIS off the deuteron~\cite{Atwood:1972zp,Landshoff:1977pg,  
Kusno:1979dk, Bodek:1980ar, Dunne:1985cn, Frankfurt:1988nt, Kaptar:1991hx, Nakano:1991kh,
Epele:1992np, Braun:1993nh, Gomez:1993ri, Melnitchouk:1994rv, Abe:1994rj, Burov:1998kz, Alekhin:2003qq,
Burov:2003yq, KP04, Kulagin:2007ph, Arrington:2008zh, Weinstein:2010rt, Arrington:2011qt}.
In spite of the broad range of predictions, such studies indicate that nuclear effects in the deuteron 
are non-negligible and rise rapidly in the region of large Bjorken $x$. 
A recent direct measurement of nuclear effects in the deuteron~\cite{Griffioen:2015hxa}
indicates a few-percent negative correction at $x\sim 0.5$-$0.6$, with a steep rise at large $x$. 
Therefore, if neglected or treated incorrectly, these nuclear effects can potentially 
introduce significant uncertainties and/or biases in the extraction of the neutron structure functions and 
of the $d$ quark distribution from the DIS data~\cite{Accardi:2011fa}.

A microscopic model for nuclear structure functions and PDFs 
accounting for a number of different nuclear effects was developed in Refs.\cite{KP04,KP07,KP10,KP14}. 
It includes the smearing with the energy-momentum distribution 
of bound nucleons (Fermi motion and binding, FMB), the off-shell correction (OS) to bound nucleon structure functions, 
the contributions from meson exchange currents and the propagation of the hadronic component of the 
virtual intermediate boson in the nuclear environment.
This model has been successfully used to quantitatively explain the 
observed $x$, $Q^2$ and $A$ dependence of 
the nuclear DIS data in a wide range of targets from ${}^3$He to ${}^{207}$Pb~\cite{KP04,KP07,KP10}, 
the magnitude, the $x$ and mass dependence of the nuclear Drell-Yan (DY) data~\cite{KP14}, as well as the data on  
the differential cross sections and asymmetries for $W^\pm,Z$ production in $p+{\rm Pb}$ collisions 
at the LHC~\cite{Ru:2016wfx}.  

A consistent description of the scattering off bound nucleons not only involves  
the smearing due to the nuclear momentum distribution, but also requires the knowledge of 
the off-shell (OS) scattering amplitudes.
The model of Ref.~\cite{KP04} exploits the observation that the nucleus is a weakly bound system and thus
it is sufficient to evaluate the OS correction to the bound nucleon PDFs in the vicinity of the mass shell.
The shape of this correction is defined by a universal function $\delta f(x)$ of the Bjorken variable $x$, 
while its nuclear dependence is driven by the average virtuality of the nucleon (off-shellness) inside 
the nucleus. 
The OS function $\delta f$ can be regarded as a special nucleon structure function, which 
does not contribute to the cross section of the physical nucleon, 
but is relevant only for the bound nucleon and describes its
response to the interaction with the nuclear environment.
The off-shell correction proved to be an important contribution to explain nuclear effects
at large $x$. The function $\delta f$ was determined from the analysis of data on ratios of
DIS structure functions in various nuclei~\cite{KP04}. 
It was also shown that in a simple single-scale model,
in which the quark momentum distributions in the nucleon
are functions of the nucleon core radius, the observed behavior of $\delta f$ can be
interpreted in terms of an increase of the confinement radius 
of the bound nucleon in the nuclear environment~\cite{KP04}. 

The deuteron is a weakly bound state of two nucleons with peculiar attributes. 
Its dynamics is better understood than the dynamics 
of many-particle nuclei, making it an ideal benchmark tool for the study of different nuclear effects. 
However, it is also considerably different with respect to 
even a three-body nucleus like ${}^3$He. For these reasons one cannot rely on simple extrapolations 
of nuclear effects from heavy targets based upon nuclear density or atomic weight, as it is 
often assumed in phenomenological analyses. In contrast, 
the model of Ref.~\cite{KP04} suggests a unified treatment of the deuteron and heavier nuclei 
on the basis of common underlying physics mechanisms. 

In this paper we discuss an independent determination of the off-shell function $\delta f$,  
together with the proton PDFs, from a global QCD analysis of proton and deuterium data.  
In Sec.\ref{sec:model} we review the model of nuclear corrections in the deuteron, while 
in Sec.~\ref{sec:fit} we discuss the details of the data analysis. In Sec.~\ref{sec:res} we compare our 
results with the ones obtained from heavy nuclear targets and discuss the impact on the uncertainties 
related to the $d/u$ ratio from global QCD fits. We summarize our results in Sec.~\ref{sec:sum}.

\section{Model of Nuclear Corrections}
\label{sec:model}

The nuclear corrections to the inelastic structure functions involve a number of different contributions. 
For the deuteron we can write 
(for simplicity we summarize the structure function $F_2$ here) \cite{KP04,KP14}:
\begin{equation}
\label{F2D}
F_2^D = F_2^{N/D} + \delta_\mathrm{MEC} F_2^D + \delta_\mathrm{coh} F_2^D,
\end{equation}
where the first term in the right-hand side stands for the incoherent scattering off the 
bound isoscalar nucleon $N$ including the off-shell correction,
and $\delta_\mathrm{MEC} F_2^D$ and $\delta_{\rm coh} F_2^D$ are the corrections
due to nuclear meson exchange currents (MEC) and
coherent interactions of the intermediate virtual boson with nuclear target, respectively.

\subsection{Incoherent Scattering off Bound Nucleons} 
\label{sec:IA} 

The first term in \eq{F2D} dominates at $x>0.2$ and can be written as follows~\cite{KP04}: 
\begin{align}
\label{eq:IA}
\gamma^2 F_2^{N/D}(x,Q^2) &=
 \int \frac{\ud^3\bm p}{(2\pi)^3} 
    \left|\Psi_D(\bm p)\right|^2
     \left(1+\frac{\gamma p_z}{M}\right)
\left({\gamma'}^2 +\frac{6{x'}^2 \bm{p}_\perp^2}{Q^2} \right)
F_2^N(x',Q^2,p^2),
\end{align}
where the integration is over the momentum of the bound nucleon $\bm p$, $\Psi_D(\bm p)$ is the deuteron wave function,
$M=\tfrac12(M_p+M_n)$ and $F_2^{\smash{N}}=\tfrac12(F_2^p+F_2^n)$ are respectively the mass and
the structure function of the bound nucleon with four-momentum $p=(M+\ceps,\bm{p})$,
where $\ceps=\ceps_D-\bm p^2/(2M)$ and $\ceps_D=M_D-2M$ is the deuteron binding energy.
The integration in \eq{eq:IA} requires the structure function of the bound nucleon
in the off-shell region and  $F_2^{\smash{N}}$ depends on the Bjorken variable $x'=Q^2/(2p q)$, 
the momentum transfer square $Q^2$,  
and also on the nucleon invariant mass squared $p^2$. 
In \eq{eq:IA} we use a coordinate system such that the momentum transfer $\bm{q}$
is antiparallel to the $z$ axis, $\bm{p}_\perp$ is the transverse component of the nucleon momentum,
and $\gamma^2=1+4x^2 M^2/Q^2$ and ${\gamma'}^2=1+4{x'}^2 p^2/Q^2$.

\begin{figure}[htb] 
\begin{center}
\includegraphics[width=0.80\textwidth]{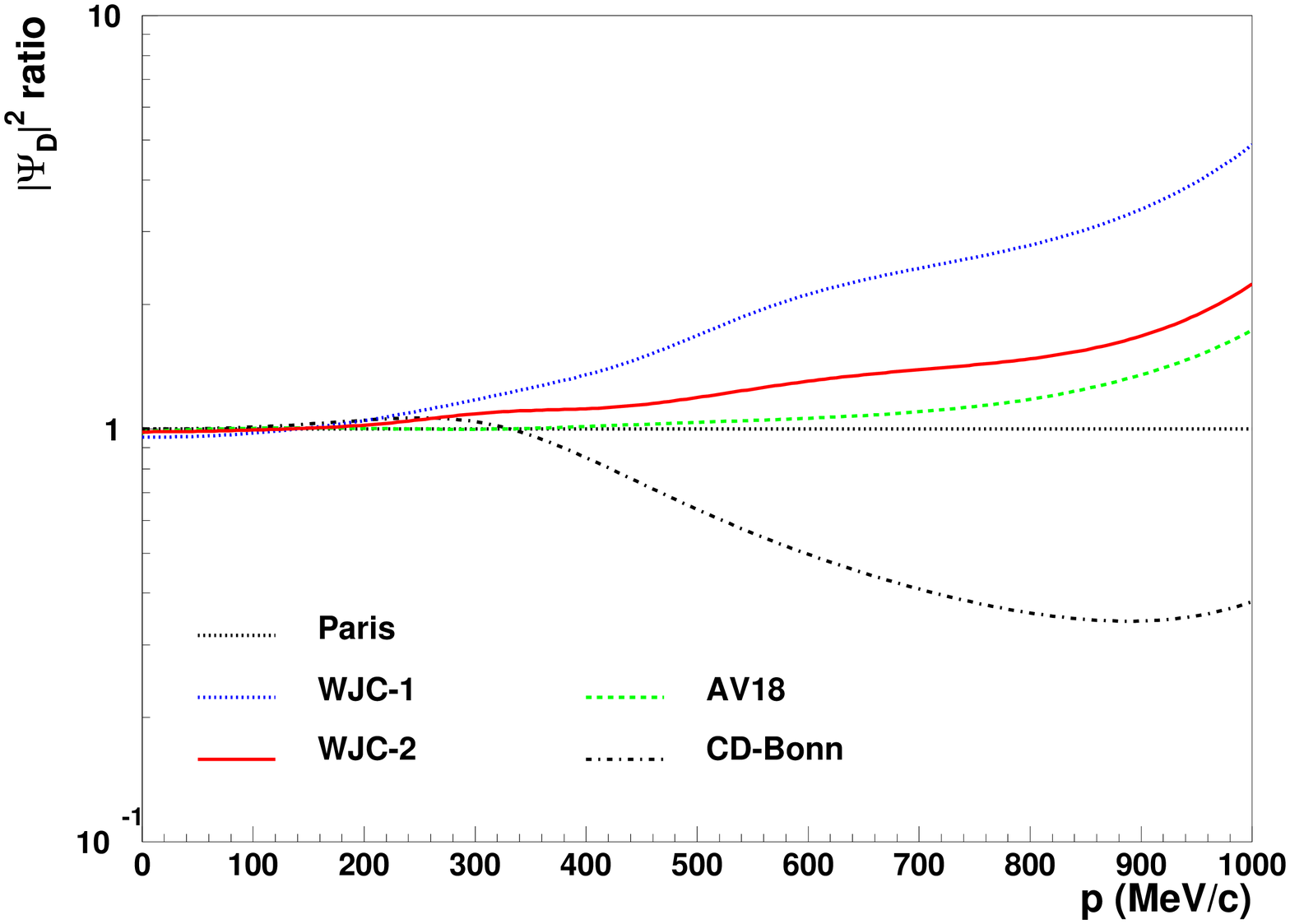}
\caption{%
Ratio of the deuteron momentum distribution $\left|\Psi_D (\bm p) \right|^2$ calculated according to various models with respect to the corresponding value in 
the Paris model. See the text for details.  
}
\label{fig:deutwf}
\end{center}
\end{figure}

The integrand in \eq{eq:IA} factorizes into two independent terms involving the contribution
from two different scales: 
i) the wave function $\Psi_D(\bm p)$ describing the deuteron properties in momentum space, and  
ii) the nucleon structure function $F_2^N$ describing processes at the parton level in the nucleon.
In the following we will consider several deuteron wave functions $\Psi_D(\bm p)$ 
corresponding to different models for the nucleon-nucleon potential: 
Paris~\cite{Lacombe:1980dr}, CD-Bonn~\cite{Machleidt:2000ge}, AV18~\cite{Veerasamy:2011ak}, 
WJC1 and WJC2~\cite{Gross:2008ps,Gross:2010qm}. 
These wave functions are constrained by high-precision fits to 
nucleon-nucleon scattering data at low energies. 
However, these models for $\Psi_D(\bm p)$ can differ by more
than a factor of 10 in the high momentum tail, as shown in Fig.~\ref{fig:deutwf}. 
Table~\ref{tab:kinwf} summarizes the salient kinematic parameters 
associated with each deuteron wave function.
To be consistent with the weak binding approximation of Ref.\cite{KP04}, 
we perform the integration over the nucleon momentum in \eq{eq:IA} up to $|\bm p|<1$\:GeV/$c$.

\begin{table}[ht]
\begin{center}
\begin{tabular}{c|c|c|c|c} \hline
Wave function  &  Reference  &  $~~~\average{v}~~~$   &  $~~~\average{\varepsilon}$~[MeV] &  
$~~~\average{\bm{p}^2}/2M$~[MeV]  \\
\hline\hline
WJC1   &  \cite{Gross:2008ps,Gross:2010qm}  & -0.062   & -16.16 & 13.94   \\  
WJC2   &  \cite{Gross:2008ps,Gross:2010qm} &  -0.049  & -12.89 &  10.67  \\ 
AV18    &  \cite{Veerasamy:2011ak}  & -0.045  & -11.93 & ~9.71  \\
Paris   &  \cite{Lacombe:1980dr}  &  -0.043   & -11.55 & ~9.33   \\
CD-Bonn    &  \cite{Machleidt:2000ge} &  -0.037 & ~-9.96 & ~7.74  \\  \hline 
\end{tabular}
\end{center}
\caption{\label{tab:kinwf}
Values of the average nucleon virtuality $v=(p^2-M^2)/M^2$, bound nucleon energy $\varepsilon$, 
and kinetic energy $\bm{p}^2/2M$ for each deuteron wave function shown in Fig.~\ref{fig:deutwf}.  
}
\end{table}

The nucleon structure function in \eq{eq:IA} includes  
the target mass and the high-twist (HT) corrections represented as follows
\begin{equation}\label{eq:SF}
F_2^N(x,Q^2,p^2) = F_2^{\text{\rm TMC}}(x,Q^2,p^2)
        + H_2^{(4)}(x)/Q^{2} + {\mathcal{O}}(Q^{-4}) 
\end{equation}
where $F_2^{\text{TMC}}$ is the leading-twist (LT) structure function
corrected for the target mass effects (TMC) and $H_2^{(4)}$ describes the
twist-4 contribution (we suppress any explicit notation to higher order terms for brevity).
The LT structure function is computed using the proton and the neutron PDFs extracted from the 
analysis of data as described in Sec.\ref{sec:fit}.
The target mass correction is computed following the prescription of Ref.~\cite{Georgi:1976ve}:
\begin{equation}\label{eq:TMC}
F_2^{\text{\rm TMC}}(x,Q^2,p^2) = \frac{x^2}{\xi^2 \gamma^3} F_2^{\rm LT}(\xi,Q^2,p^2) + 
\frac{6x^3p^2}{Q^2\gamma^4}\int^1_\xi \frac{dz}{z^2} F_2^{\rm LT}(z,Q^2,p^2) + {\mathcal{O}}(Q^{-4}) 
\end{equation}
where $\xi=2x/(1+\gamma)$ is the Nachtmann variable, we substitute $p^2$ for the
mass of the bound nucleon squared $M^2$, and $\gamma^2=1+4x^2 p^2/Q^2$. 
Note that the second term in \eq{eq:TMC} is suppressed as $1/Q^2$ and 
therefore can be formally considered as a kinematic HT contribution. 
Recent phenomenology suggests that \eq{eq:SF} with twist-4 contributions 
provides a good description of data down to $Q\sim 1\gev$
\cite{Alekhin:2007fh}.
It is also worth noting that this model is consistent with duality principle and on average
describes the resonance data with $W<1.8\gev$ \cite{Alekhin:2007fh,Malace:2009kw}.

\subsection{Off-shell Correction} 
\label{sec:OS} 

The structure function of the bound nucleon $F_2^N(x,Q^2,p^2)$ appearing in the calculation of 
the nuclear correction in \eq{eq:IA} explicitly depends on the nucleon invariant 
mass squared $p^2$. 
The $p^2$-dependence of the structure function has two different sources \cite{Kulagin:1994fz,KP04}:
(i) the dynamic off-shell dependence of the LT structure function; 
(ii) the kinematic target mass correction, which generates terms of the order of $p^2/Q^2$. 
We evaluate the off-shell dependence of the target mass correction by
replacing $M^2\to p^2$ in \eq{eq:TMC}.
Since the characteristic momenta of a bound nucleon are small compared to its mass,
the integration in \eq{eq:IA} mainly covers a region in the vicinity of the mass shell. 
The nucleon virtuality $v=(p^2-M^2)/M^2$ can then be treated as a small parameter, so that we  
can expand the structure function in series of $v$ keeping only the leading term:
\begin{align}
\label{SF:OS}
F_2^{\rm LT}(x,Q^2,p^2) &=
F_2^{\rm LT}(x,Q^2)\left[ 1+\delta f(x,Q^2)\,v \right],\\
\label{delf}
\delta f &= \partial\ln F_2^{\rm LT}/\partial\ln p^2\mid_{p^2=M^2},
\end{align}
where the first term on the right-hand side in \eq{SF:OS} is the structure
function of the on-mass-shell nucleon.
The off-shell (OS) function $\delta f$ can be regarded as a special nucleon structure function,
which describes the relative  
modification of the nucleon structure functions and PDFs in the vicinity of the mass shell.
This function does not contribute to the cross section of the
physical nucleon, but it is relevant only for the bound nucleon and describes its
response to the interaction in a nucleus.
  
In general, the function $\delta f$ might be flavor dependent and different
for protons and neutrons.
However, the study of nuclear DIS and DY data~\cite{KP04,KP10,KP14} 
supports the hypothesis of the OS function universality, 
with no significant $Q^2$ dependence suggested by the data, i.e. $\delta f(x,Q^2)=\delta f(x)$. 
Although we assume that $\delta f$ is only a function of $x$, the overall OS correction to the nuclear structure 
functions also depends on $Q^2$, as a result of the integration of \eq{eq:IA}.
It is important to note that this $Q^2$ dependence is different from the ones of both 
the LT and HT contributions to the structure functions in \eq{eq:SF}. 
This difference allows a simultaneous extraction of PDFs, HTs and $\delta f$ from 
global QCD fits (cf. Sec.~\ref{sec:fit}).

\begin{figure}[htb] 
\begin{center}
\includegraphics[width=0.8\textwidth]{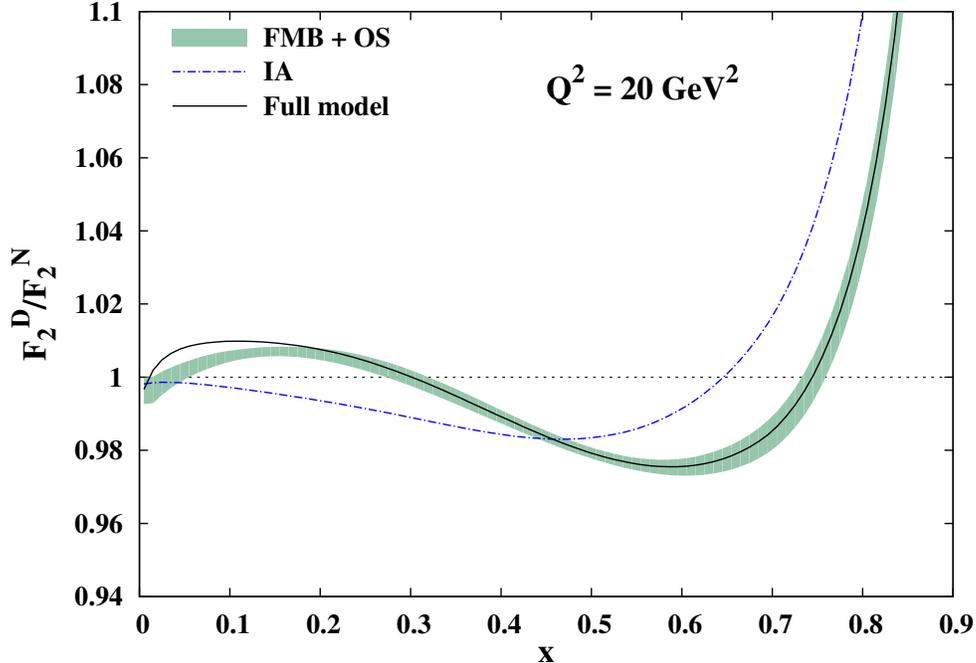}
\caption{%
Ratio of the deuteron and the isoscalar nucleon structure functions $F_2^D/F_2^N$ calculated at $Q^2=20\gevsq$
using different approximations. The solid line is the full model of Ref.\cite{KP04}, 
while the dashed line is the result of Ref.\cite{KP04} with no off-shell ($p^2=M^2$), 
nuclear pion and nuclear shadowing correction (impulse approximation). The shaded area represents the 
$\pm 1 \sigma$ uncertainty band on the impulse approximation supplemented by the off-shell correction only.
}
\label{fig:d_n}
\end{center}
\end{figure}

\subsection{Other Nuclear Corrections} 
\label{sec:others} 


The remaining corrections appearing in \eq{F2D}, i.e. the nuclear meson exchange current 
$\delta_\mathrm{MEC} F_2^D$ and the nuclear shadowing (NS) $\delta_{\rm coh} F_2^D$,  
are relevant only at small $x$.
For the details of the treatment of these terms we refer to Ref.\cite{KP04}.
Here we only emphasize that nuclear effects present in different kinematical regions of
$x$ are related by the DIS sum rules and normalization constraints~\cite{KP14}.
For example, the light-cone momentum sum rule links the FMB and the MEC corrections. 
We use this relation to constrain the mesonic contributions to the nuclear structure functions. 
Similarly, the baryon number sum rule links the shadowing and the OS corrections. 
In our approach, the OS effect provides the mechanism to cancel a negative nuclear-shadowing 
contribution to the normalization of the nuclear valence quark distributions.

\subsection{Model Predictions and Phenomenology} 
\label{sec:phen} 


The model described above was used to perform a detailed analysis of data on 
the ratio $\mathcal R(A/B) = F_2^A/F_2^B$ between the nuclear targets $A$ and $B$~\cite{KP04}.  
The data sets analyzed stem from a variety of electron and muon DIS experiments 
(CERN NMC, EMC, and BCDMS, SLAC E139 and E140, Fermilab E665), with targets ranging from $^4$He to $^{208}$Pb 
in a wide region of $x$ and $Q^2$.
In this way we tested the hypothesis that the OS modification of bound nucleons is responsible for 
the difference between the data and all known nuclear effects in \eq{F2D}, including the 
FMB~\cite{Akulinichev:1985ij,Kulagin:1989mu}, the nuclear shadowing and the nuclear MEC.  
In turn, this OS correction is controlled by the universal off-shell function $\delta f$ in \eq{SF:OS}, 
which was determined from DIS data with the corresponding uncertainty~\cite{KP04}.  

Such an approach leads to an excellent agreement with the available DIS data on 
the $x$, $Q^2$, and $A$ dependence of $\mathcal R(A/B)$. 
The model predictions are also in a good agreement \cite{KP10} 
with the recent DIS data by the HERMES experiment at HERA \cite{Ackerstaff:1999ac}
and the E03-103 experiment at JLab~\cite{Seely:2009gt} down to ${}^3$He.
Furthermore, the same model allows a calculation of nuclear PDFs, which can describe well the 
magnitude, the $x$ and mass dependence of the available data on Drell-Yan production 
off various nuclear targets~\cite{KP14}, as well as the differential cross sections 
and asymmetries for $W^\pm,Z$ production in $p+{\rm Pb}$ collisions at the LHC~\cite{Ru:2016wfx}.  


In this paper we perform an independent analysis of 
deuterium and proton data in the context of global QCD fit.
In Fig.\ref{fig:d_n} we show the predictions of Ref.~\cite{KP04} for the ratio 
$\mathcal{R}(D/N)=F_2^D/F_2^N$    
of the deuteron and the isoscalar nucleon structure functions at $Q^2=20 \gevsq$.
The region of $0.35 \leq x \leq 0.55$ is characterized by an almost linear dependence from $x$, with a slope  
$\ud {\mathcal R}(D/N) / \ud x = - 0.099\pm 0.006$, including model uncertainties. 
This slope is often used in the analysis 
of experimental data~\cite{Seely:2009gt} since it is less affected by the experimental uncertainties 
(especially the overall normalization) than the absolute value of the nuclear correction.  
At large $x>0.1$ nuclear corrections are dominated by the FMB and OS effects, as shown in Fig.~\ref{fig:d_n}. 
In particular, the off-shell correction is a crucial contribution in this kinematic region, which is 
studied in more detail in the present analysis.

\begin{table}[ht] 
\begin{center}
\begin{tabular}{l|l|c|c|c|c|c} \hline 
            & Experiment & Reference & Beam & Target(s) & Final states & Data points \\ 
\hline\hline
DIS collider      &  HERA I+II & \cite{Abramowicz:2015mha}  & $e$ & $p$ & $e X$ & 1168 \\  
                  &                  &     &  &  & $\nu X$ &  \\  
                  &  HERA I+II  & \cite{Abramowicz:1900rp}  & $e$ & $p$ & $e c X$ & 52 \\   
                  &  H1  & \cite{Aaron:2009af}  & $e$ & $p$ & $e b X$ & 12 \\   
                  &  ZEUS  & \cite{Abramowicz:2014zub}  & $e$ & $p$ & $e b X$ & 17 \\ \hline  
DIS fixed target  & BCDMS & \cite{Benvenuti:1989rh,Benvenuti:1989fm} & $\mu$ & $p,D$ & $\mu X$ & 605 \\  
                  & NMC & \cite{Arneodo:1996qe} & $\mu$ & $p,D$ & $\mu X$ & 490 \\  
                  & SLAC E49a & \cite{Bodek:1979rx} & $e$ & $p,D$ & $e X$ & 118 \\  
                  & SLAC E49b & \cite{Bodek:1979rx} & $e$ & $p,D$ & $e X$ & 299 \\  
                  & SLAC E87 & \cite{Bodek:1979rx} & $e$ & $p,D$ & $e X$ & 218 \\  
                  & SLAC E89b & \cite{Mestayer:1982ba} & $e$ & $p,D$ & $e X$ & 162 \\  
                  & SLAC E139 & \cite{Gomez:1993ri} & $e$ & $D$ & $e X$ & 17 \\  
                  & SLAC E140 & \cite{Dasu:1993vk} & $e$ & $D$ & $e X$ & 26 \\  
                  & JLab BONuS & \cite{Griffioen:2015hxa} &  $e$ & $D$ & $eX$ & 5 \\ 
                  & NOMAD & \cite{Samoylov:2013xoa} & $\nu$ & {\it Fe} & $\mu^+ \mu^- X$ & 48 \\ 
                  & CHORUS & \cite{KayisTopaksu:2011mx} & $\nu$ & {\it Emul.} & $\mu c X$ & 6 \\   
                  & CCFR & \cite{Goncharov:2001qe} & $\nu$ & {\it Fe} & $\mu^+ \mu^- X$ & 89 \\ 
                  & NuTeV & \cite{Goncharov:2001qe} & $\nu$ & {\it Fe} & $\mu^+ \mu^- X$ & 89 \\ \hline  
Drell-Yan fixed target  & FNAL E866 & \cite{Towell:2001nh} & $p$ & $p,D$ & $\mu^+ \mu^-$ & 39 \\  
                        & FNAL E605 & \cite{Moreno:1990sf} & $p$ & {\it Cu} & $\mu^+ \mu^-$ & 119 \\ \hline 
$W,Z$ collider    & D0 & \cite{Abazov:2013rja} & $p$ & $\bar p$ & $W^+\to \mu^+\nu$  & 10 \\  
    &  &  &  &  & $W^-\to \mu^-\nu$  &  \\  
                  & D0 & \cite{D0:2014kma} & $p$ & $\bar p$ & $W^+\to e^+\nu$  & 13 \\ 
    &  &  &  &  & $W^-\to e^-\nu$  &  \\  
                  & ATLAS & \cite{Aad:2011dm,Aad:2016naf} & $p$ & $p$ & $W^+\to l^+\nu$  & 36 \\ 
    &  &  &  &  & $W^-\to l^-\nu$  &  \\  
    &  &  &  &  & $Z\to l^+l^-$  &  \\  
                  & CMS & \cite{Chatrchyan:2013mza,Khachatryan:2016pev} & $p$ & $p$ & $W^+\to \mu^+\nu$  & 33 \\ 
    &  &  &  &  & $W^-\to \mu^-\nu$  &  \\  
                  & LHCb & \cite{Aaij:2015gna,Aaij:2015zlq} & $p$ & $p$ & $W^+\to \mu^+\nu$  & 63 \\ 
    &  &  &  &  & $W^-\to \mu^-\nu$  &  \\  
    &  &  &  &  & $Z\to \mu^+\mu^-$  &  \\  
                  & LHCb & \cite{Aaij:2015vua} & $p$ & $p$ & $Z\to e^+e^-$  & 17 \\ \hline  
\end{tabular}
\end{center}
\caption{\label{tab:data}
List of the various data sets used in the present analysis. 
}
\end{table}

\section{Off-shell Correction from Global QCD Fit} 
\label{sec:fit}

In this paper we discuss the impact of nuclear effects in the deuteron in the context of global QCD fits. 
Our goals are twofold: (i) an independent determination of the off-shell correction preferred by 
the deuteron data; 
(ii) an estimate of the PDF uncertainties (in particular for the $d/u$ ratio at large $x$) introduced by the 
nuclear corrections to deuterium data.  
The analysis framework and the main data sets used are common to the ABMP16 fit~\cite{Alekhin:2017kpj}.

\subsection{Analysis Framework} 
\label{sec:framework} 

In our analysis we use the next-to-next-to-leading-order (NNLO) approximation in the QCD perturbation theory 
to calculate the partonic cross sections entering the LT terms for the hard interaction processes considered. 
We set the renormalization and factorization scales to $\mu_r=\mu_f=\mu$ and we identify this scale $\mu$ with the 
relevant kinematics of each process, e.g. $\mu = Q$ for DIS. 
The individual PDFs are parametrized as in Ref.~\cite{Alekhin:2017kpj} 
at the starting scale $\mu^2=Q^2_0=9$ GeV$^2$. The PDFs are subject to sum rule 
constraints due to the conservation of the quark number and the momentum in the nucleon. 

The splitting functions controlling the scale dependence of the PDFs in the evolution equations are 
evaluated at NNLO in perturbation theory~\cite{Moch:2004pa,Vogt:2004mw}. 
The Wilson coefficients entering the massless DIS structure functions are calculated at  
NNLO~\cite{vanNeerven:1991nn,Zijlstra:1991qc,Zijlstra:1992qd,Zijlstra:1992kj,Moch:1999eb,Moch:2004xu,Vermaseren:2005qc}.
Similarly, we use NNLO calculations for the partonic cross sections of the Drell-Yan process and the 
hadronic $W$ and $Z$ boson production~\cite{Hamberg:1990np,Harlander:2002wh,Anastasiou:2003yy,Anastasiou:2003ds,Catani:2009sm}.  

In our PDF analysis we use a fixed flavor number scheme with $n_f=3$ light flavors from which heavy quark PDFs 
are generated. The heavy quark masses $m_q$ are defined in the $\overline{\rm MS}$ renormalization scheme  
as running masses $m_q(\mu)$ depending upon the scale $\mu$ of the hard scattering in analogy to the running 
coupling $\alpha_s(\mu)$. 
As discussed in Ref.~\cite{Alekhin:2010sv}, the use of the $\overline{\rm MS}$-mass allows better 
convergence properties and greater perturbative stability at higher orders. 
The heavy quark Wilson coefficients entering the DIS structure functions for heavy quark production 
are known exactly only to the next-to-leading-order (NLO) for both the 
charged current (CC)~\cite{Gottschalk:1980rv,Gluck:1996ve} and neutral current (NC)~\cite{Laenen:1992zk} processes.   
For the NC case approximate NNLO coefficients are employed~\cite{Alekhin:2017kpj}. 

In the kinematic range of our analysis the twist-6 terms give negligible contributions to 
the structure functions~\cite{Alekhin:2007fh}.   
Therefore, in addition to the leading twist we only include two twist-4 contributions 
$H_2$ and $H_T$ -- as defined in \eq{eq:SF} -- to the structure functions $F_2$ and $F_T$, respectively. 
We also considered the target dependence of the HT parametrization. The isospin asymmetry 
in $H_T$ is consistent with zero within uncertainties~\cite{Alekhin:2003qq} and therefore 
is neglected in our analysis. 
The isospin asymmetry in $H_2$ is also small~\cite{Alekhin:2003qq}. 
Although the values of this latter have a better statistical significance, we set it to zero as well 
in order to avoid a potential bias in the nuclear corrections extracted from 
the global QCD fits.  
In summary, we fit two twist-4 coefficients for the isoscalar nucleon, $H_2^N$ and $H_T^N$. 
These power corrections are parametrized as cubic spline functions of $x$. 
 
The nuclear corrections for the deuteron are calculated according to the model described in 
Sec.~\ref{sec:model}. We do not include meson exchange currents and coherent nuclear effects
(shadowing) for the deuteron, since their impact 
is negligible in the kinematic coverage of our analysis (see Fig.~\ref{fig:d_n}) and we 
are mainly focused on the study of the off-shell correction. 
The only free parameters entering the nuclear corrections are the ones describing the 
off-shell function $\delta f(x)$, which are extracted simultaneously with the PDFs and 
HT terms. To this end, we use a parametrization with generic second- and third-order polynomials 
for $\delta f(x)$. We verified that there is no statistically significant difference between these two 
options within the accuracy of the data samples used in our analysis.

\begin{figure}[htb] 
\begin{center}
\includegraphics[width=1.0\textwidth]{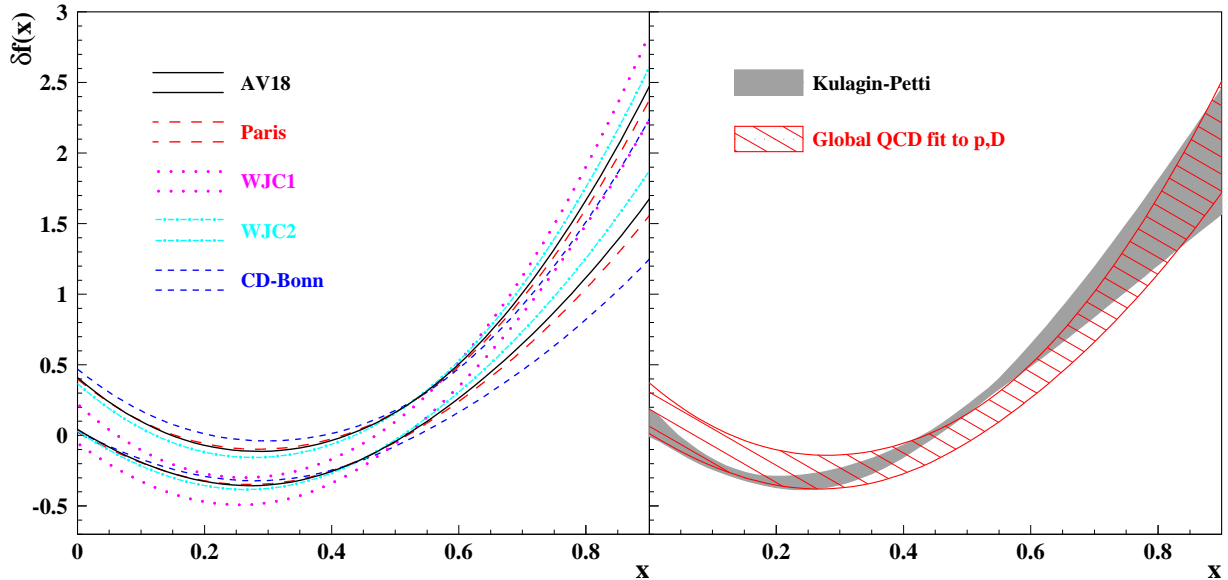}
\caption{%
Left panel: Comparison of the off-shell functions $\delta f(x)$ ($\pm 1 \sigma$ uncertainty bands) 
extracted within our global QCD fit including all data sets in Table~\ref{tab:data} by 
using different models for the deuteron wave functions. 
Right panel: Summary $\pm 1 \sigma$ uncertainty band on $\delta f(x)$ obtained from this analysis,   
including the statistical uncertainty from the fit and the one related 
to the choice of the deuteron wave function. 
The corresponding uncertainty band obtained from heavy target data ($A \geq 4$) in Ref.~\cite{KP04} is 
shown for comparison. 
}
\label{fig:dfwf}
\end{center}
\end{figure}

\subsection{Data Samples} 
\label{sec:data} 

In our analysis most of the information about the deuteron is provided by the 
inclusive DIS data off deuterium from the SLAC E49, E87, E89, E139, 
E140~\cite{Bodek:1979rx,Atwood:1976ys,Mestayer:1982ba,Gomez:1993ri,Dasu:1993vk} 
and the CERN BCDMS~\cite{Benvenuti:1989fm}, NMC~\cite{Arneodo:1996qe} 
experiments, as well as by the ratio of Drell-Yan production 
in $p{\rm D}$ and $pp$ collisions from the Fermilab E866 experiment~\cite{Towell:2001nh}.  
In addition, the recent direct measurement of the ratio $F_2^D/F_2^N$~\cite{Griffioen:2015hxa} 
by the BONuS experiment~\cite{Tkachenko:2014byy} 
at Jefferson laboratory allows a better disentanglement of the nuclear corrections in the deuteron 
from possible variations of the $d/u$ ratio in the nucleon. Since most of the BONuS data 
either have low values of $Q^2$ or are in the resonance region, 
we only include the BONuS points with $Q^2>1.5$ GeV$^2$ and $W>1.6$ GeV.  
Although these cuts are less stringent than the ones we apply for the other data sets, they 
are justified by a partial cancellation of HT effects in the ratio $F_2^D/F_2^N$ and  
by the relevance of the direct BONuS measurement for our study. 

For consistency with the ABMP16 fit~\cite{Alekhin:2017kpj} and to better constrain the sea 
quark distributions we include the Drell-Yan data in $p{\rm Cu}$ interactions by the E605 experiment, 
as well as charm production data in (anti)neutrino interactions off heavy targets by the  
CCFR~\cite{Goncharov:2001qe}, NuTeV~\cite{Goncharov:2001qe}, NOMAD~\cite{Samoylov:2013xoa}, 
and CHORUS~\cite{KayisTopaksu:2011mx} experiments. We verified that these data using 
nuclear targets do not affect our results on the deuteron by performing dedicated fits with and without 
such data. It is also worth noting that NOMAD and CHORUS measured the ratios of 
charm to inclusive charged-current cross sections and it was shown that the 
corresponding nuclear corrections cancel out at the subpercent level in such 
ratios~\cite{Alekhin:2014sya}. 

Nuclear corrections in the deuteron can be determined by comparing the available data for deuterium targets 
with the ones originated from interactions on the free nucleons. 
However, inclusive DIS data off protons do not allow to disentangle the $d$ and $u$ quark distributions 
because of the lack of a corresponding free neutron target. 

The limitations of DIS data can be partially overcome with the addition of Drell-Yan, $W^\pm$, and $Z$ production 
at Tevatron and the LHC~\cite{Alekhin:2017kpj,Abazov:2013rja,D0:2014kma,Aad:2011dm,Aad:2016naf,Chatrchyan:2013mza,Khachatryan:2016pev,Aaij:2015gna,Aaij:2015vua,Aaij:2015zlq}. 
In particular, the data on $W^+$ and $W^-$ 
production allow a $d/u$ separation independent from the deuterium data.  
We note that the same $W^\pm$ production data sets collected at Tevatron and the LHC may result 
in two distinct (but correlated) measurements: (i) the $l^\pm$ lepton asymmetry from the $W^\pm$ decays;  
(ii) the actual $W^\pm$ asymmetry. The former is more closely related to the experimental 
observables, while the latter requires model-dependent acceptance corrections to account 
for the kinematics of the $W^\pm$ decay. As discussed in Ref.~\cite{Alekhin:2015cza},  
some inconsistencies between the $l^\pm$ lepton asymmetries and the $W^\pm$ asymmetries obtained 
from the same experimental data sets are observed. For this reason whenever both measurements are available, 
we only consider the $l^\pm$ lepton asymmetry data in our analysis.  

Table~\ref{tab:data} summarizes all the data sets used in the present analysis. 
In order to exclude the region of resonance production and to reduce the impact of HT corrections 
we require $Q^2>2.5$ GeV$^2$ and $W>1.8$ GeV for DIS data.

\section{Results and Discussion}
\label{sec:res}

The general features of the PDFs extracted from the global fits described above as well as a detailed 
discussion of the individual data samples considered were presented elsewhere~\cite{Alekhin:2017kpj}.
In this paper we focus on the nuclear correction extracted from deuterium data and on the corresponding 
impact for the $d/u$ ratio.

\subsection{Off-shell Function $\delta f$}
\label{sec:deltaf}

The results of our determination of the off-shell function $\delta f(x)$ from the global 
QCD fits described in Sec.~\ref{sec:fit} are shown in Fig.~\ref{fig:dfwf}. A simultaneous 
extraction of the off-shell function with both PDFs and HTs is possible because of the different 
$Q^2$ dependence of these three contributions and the wide $Q^2$ coverage of the data 
sets listed in Table~\ref{tab:data}. In general, nuclear corrections to the deuteron data are 
partially correlated to the $d$-quark distribution. In order to reduce this correlation, the role of 
Drell-Yan and $W^\pm$ production at $pp$ and $p\bar{p}$ colliders is crucial. In particular, the 
recent combined D0 data and the LHC data from LHCb, reaching values of $x\sim 0.8$ 
due to the wide rapidity coverage, offer precisions comparable to the ones of older fixed-target 
DIS experiments. 

Figure~\ref{fig:dfwf} illustrates the dependence of the fitted $\delta f(x)$ function upon the 
choice of the deuteron wave function $\Psi_D(\bm p)$ among the models listed in Sec.~\ref{sec:IA}. 
The main differences are related to the high momentum component of the
wave function, as shown in Fig.~\ref{fig:deutwf}.
Since this high momentum tail controls the region of large nucleon virtuality $v$, 
the off-shell correction in the large $x$ region is in principle sensitive to the 
corresponding nuclear smearing in \eq{eq:IA}, which modifies the $x$ and $Q^2$ 
dependence of the structure functions.  
A general trend can be observed from Fig.~\ref{fig:dfwf}, with the harder 
wave function resulting in a slightly higher off-shell function at large $x$. 
Since the overall off-shell correction has opposite sign with respect to $\delta f(x)$ 
in \eq{SF:OS}, this trend implies an anticorrelation between $\Psi_D(\bm p)$ 
and $v\delta f(x)$ in global QCD fits.   
From Fig.~\ref{fig:dfwf} we note that our results obtained with the Paris, CD-Bonn, 
AV18, WJC1, and WJC2 wave functions are all consistent within the corresponding uncertainties 
and indicate a relatively limited spread. This robustness against the  modeling 
of the deuteron wave function can be explained by the use of data samples which can 
reduce the correlation between the nuclear correction and the $d$-quark distribution. 
In this context the recent direct measurement of the ratio $F_2^D/F_2^N$ from the 
BONuS experiment contributes to constrain the overall normalization 
of the nuclear corrections in our fits.

\begin{figure}[htb] 
\begin{center}
\includegraphics[width=0.80\textwidth]{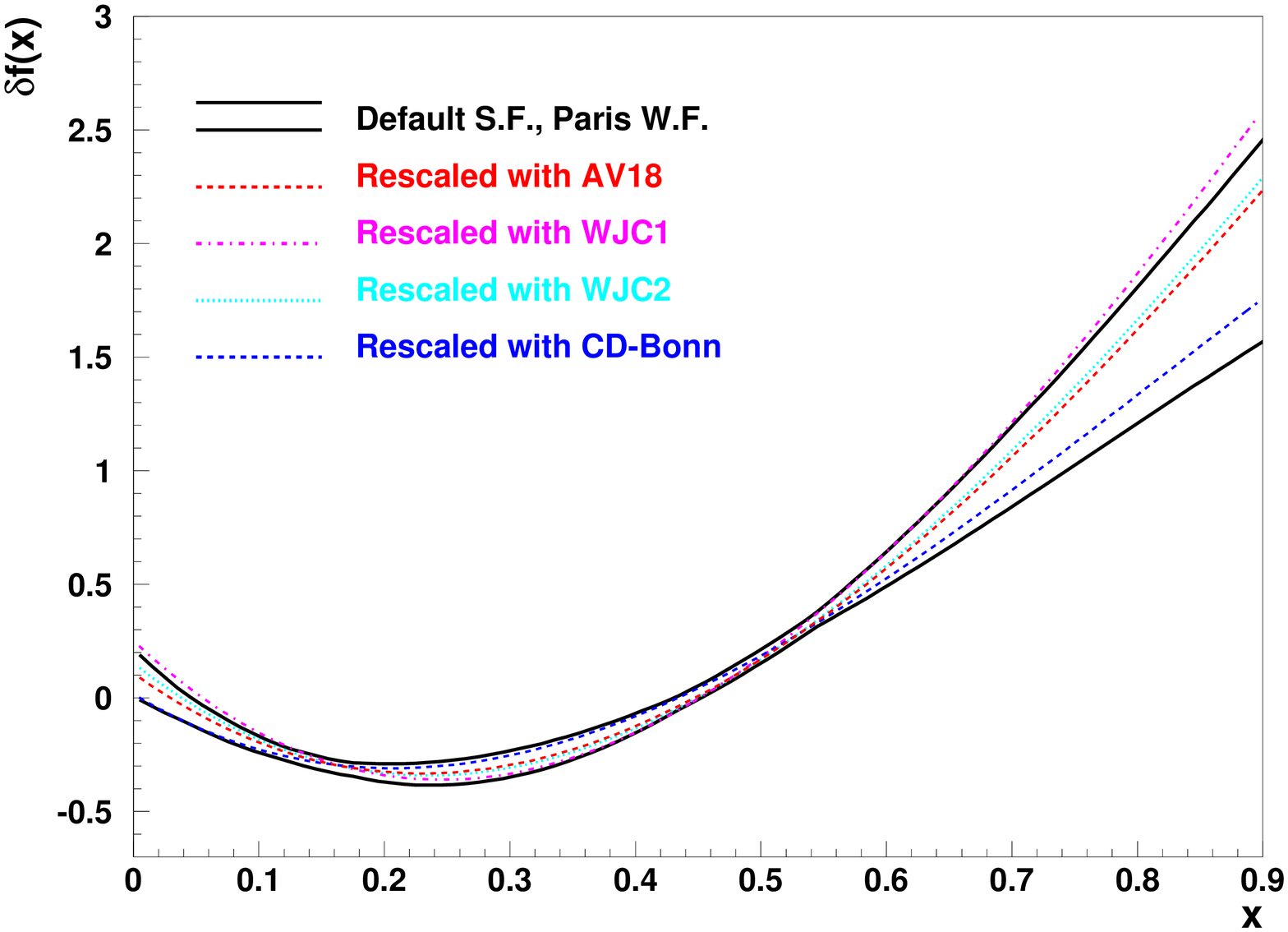}
\caption{%
Comparison of the off-shell functions $\delta f(x)$ 
extracted from the analysis of the ratios of nuclear structure functions with $A \geq 4$~\cite{KP04} by
varying the nuclear spectral function and the deuteron wave function. The nuclear spectral function 
has been rescaled by the ratios of the various models for the deuteron wave functions shown 
in Fig.~\ref{fig:deutwf}. 
The solid band represents the overall $\pm 1 \sigma$ uncertainty 
on $\delta f(x)$ from Ref.~\cite{KP04}, including model systematics. 
}
\label{fig:dfsf}
\end{center}
\end{figure}

A more precise determination of the off-shell function $\delta f(x)$ was obtained in Ref.~\cite{KP04} 
from heavy nuclei with $A\geq 4$, as described in Sec.~\ref{sec:phen}. In order to further study 
the sensitivity to the nuclear smearing in \eq{eq:IA}, we repeat the standalone extraction of 
$\delta f(x)$ in Ref~\cite{KP04} after rescaling the nuclear spectral function describing the 
properties of heavy nuclei~\cite{KP04} by the ratios of the 
various deuteron wave functions shown in Fig.~\ref{fig:deutwf}. 
The results summarized in Fig.~\ref{fig:dfsf} demonstrate a small sensitivity to the 
choice of the nuclear spectral function and/or of the deuteron wave function, as well as 
a dramatic reduction of the uncertainties with respect to Fig.~\ref{fig:dfwf}.  
This reduction can be explained by the different observables 
considered in the two independent extractions. In the global QCD fits we use the absolute DIS 
cross sections off the deuteron, while in the standalone determination of Ref~\cite{KP04} 
we consider only {\rm ratios} $\mathcal R(A/B) = F_2^A/F_2^B$ between two nuclear targets $A$ and $B$. 
Many model uncertainties largely cancel out in such ratios. 
For the same reason the data sets used in Ref~\cite{KP04} 
are more accurate than the deuteron ones, making them an excellent tool to study 
the off-shell function $\delta f(x)$. The $\pm1 \sigma$ 
uncertainty band shown in Fig.~\ref{fig:dfsf} includes model systematics due to the  
spectral and wave functions, the functional form, the PDFs, as well as corrections due to 
meson exchange currents and nuclear shadowing. 

A comparison between the two independent determinations of the off-shell function $\delta f$ 
is given in Fig.~\ref{fig:dfwf}. 
Since the five individual determinations of the off-shell function $\delta f$ from 
deuterium data using different wave functions are characterized by a comparable fit quality, 
we combine them by taking an average of both the central values and the 
corresponding uncertainties. The resulting $\pm 1 \sigma$ uncertainty band is shown in the 
right panel of Fig.~\ref{fig:dfwf}.  
This band summarizes our determination of $\delta f$ from deuteron data  
and is consistent with the more precise determination of $\delta f$ from the analysis of the 
ratios of nuclear structure functions for $A\geq 4$~\cite{KP04}. 
Since we are using a generic polynomial to parametrize 
$\delta f$ (Sec.~\ref{sec:framework}), no functional form bias is present in this comparison.  
The agreement between the two independent determinations 
supports the interpretation of the off-shell function
$\delta f$ as a universal structure function of the nucleon, validating the unified treatment 
of the deuteron and heavier nuclei developed in Ref~\cite{KP04}.

\begin{figure}[tb] 
\begin{center}
\includegraphics[width=0.8\textwidth]{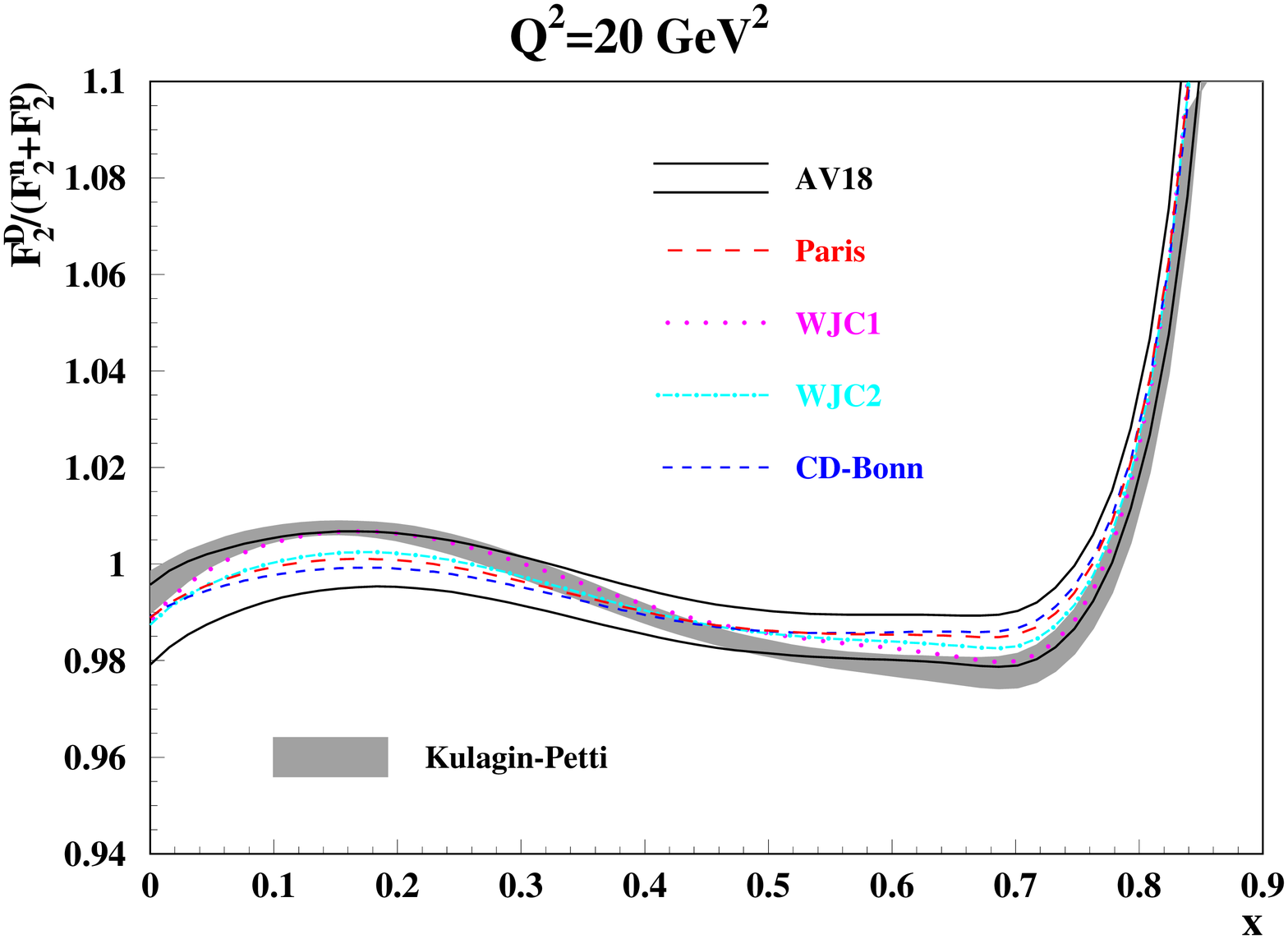}
\caption{%
Ratio $F_2^D/F_2^N$ between the deuteron and the isoscalar nucleon structure functions and its 
$\pm 1 \sigma$ uncertainty band obtained using the off-shell functions in Fig.~\ref{fig:dfwf} 
convoluted with the corresponding models for the deuteron wave function.
The $\pm 1 \sigma$ band for the same ratio obtained with the off-shell function $\delta f$ 
from Ref.~\cite{KP04} is displayed as a shaded area for comparison. 
}
\label{fig:f2d}
\end{center}
\end{figure}

\subsection{Nuclear Corrections to $F_2^D/F_2^N$}
\label{sec:f2df2n}

The nuclear correction stemming from the FMB and OS effects on the ratio $F_2^D/F_2^N$ 
is given in Fig.~\ref{fig:f2d}. This ratio is particularly interesting because it represents  
the overall nuclear corrections in the deuteron. 
The variation due to the choice of the deuteron wave function 
in the global QCD fit appears to be even smaller in the ratio $F_2^D/F_2^N$ 
than in the off-shell functions $\delta f$ given in Fig.~\ref{fig:dfwf}. This behavior  
can be explained by the anticorrelation between $\Psi_D(\bm p)$ and $v\delta f(x)$ discussed 
in Sec.~\ref{sec:deltaf}: a larger off-shell function partially compensates a 
reduced strength of the high momentum component of the wave function so that the 
observable structure function remains consistent with the fitted data.  
In addition, the recent BONuS measurement significantly constrains the ratio $F_2^D/F_2^N$, 
as mentioned in Sec.~\ref{sec:deltaf}. 
The results obtained from deuteron data with such a constraint agree 
with the predictions from Ref~\cite{KP04} based upon a standalone analysis of 
heavy nuclei with $A\geq4$, as shown in Fig.~\ref{fig:f2d}.~\footnote{ 
The result with the off-shell function $\delta f$ from Ref.~\cite{KP04} shown 
in Fig.~\ref{fig:f2d} is slightly different with respect to the calculation in Fig.~\ref{fig:d_n}. 
The differences are mainly at large $x$ values and appear due to the fact that the results 
shown in Fig.~\ref{fig:f2d} are obtained with the PDFs and HT terms extracted from our global 
QCD fit.} 

\begin{figure}[tb] 
\begin{center}
\includegraphics[width=0.8\textwidth]{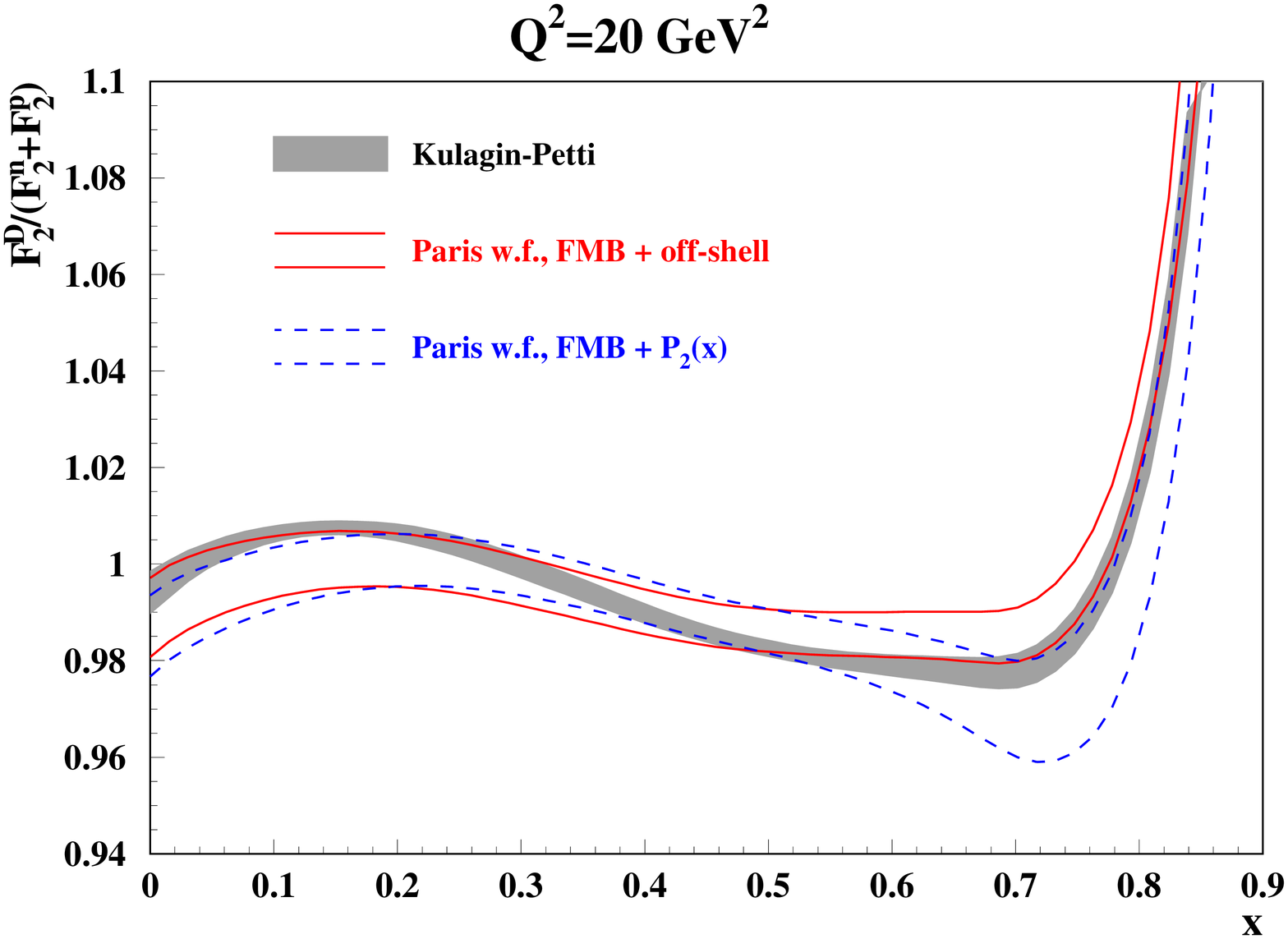}
\caption{%
Test of model dependence in the extraction of the off-shell function. The dashed lines represent 
the $\pm 1 \sigma$ uncertainty band obtained by fitting a generic polynomial as off-shell 
correction to $F_2^D/F_2^N$, instead of using the nuclear convolution with $\delta f$ 
in \eq{eq:IA} (solid lines). 
The $\pm 1 \sigma$ uncertainty band obtained from heavy target data ($A \geq 4$) in Ref.~\cite{KP04} is
displayed as shaded area for comparison.
}
\label{fig:f2d-model}
\end{center}
\end{figure}

Although the off-shell function $\delta f$ is extracted in our analysis as a generic 
polynomial, we are still calculating the nuclear correction to the structure functions 
using the nuclear convolution in \eq{eq:IA}, following the prescriptions of 
the model of Ref.~\cite{KP04}. In order to verify whether this procedure introduces 
any indirect model dependence in our results, we perform a separate fit in which we 
parametrize the overall off-shell 
correction to the structure function $F_2^D$ as a generic polynomial added 
to the standard FMB correction. In this approach the fitted off-shell correction 
is model independent as it is not part of the nuclear convolution in \eq{eq:IA}. 
The results obtained with such a parametrization shown in Fig.~\ref{fig:f2d-model} 
are in good agreement with the corresponding fits based upon the 
nuclear convolution with the off-shell function $\delta f$.  
We can thus conclude 
that the functional form we are using in our fits for $\delta f$ is flexible enough to 
reproduce the data and that our modeling of nuclear effects does not introduce any 
significant bias.

\begin{figure}[htb] 
\begin{center}
\includegraphics[width=0.8\textwidth]{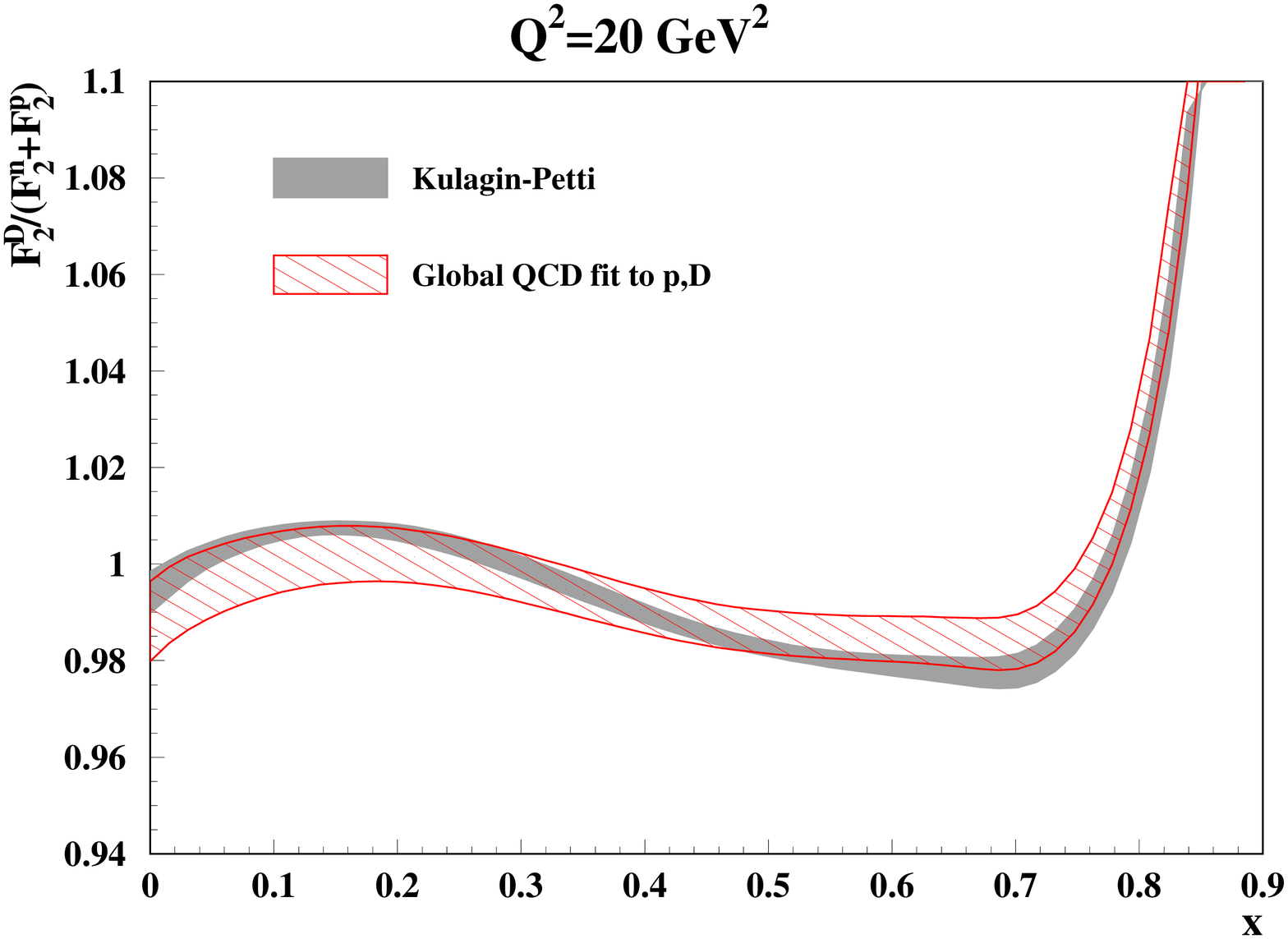}
\caption{%
Summary of the ratio $F_2^D/F_2^N$ obtained from this analysis with the corresponding 
total $\pm 1 \sigma$ band including the fit uncertainty and the one related to the choice of 
the deuteron wave function. 
The $\pm 1 \sigma$ uncertainty band obtained from heavy target data ($A \geq 4$) in Ref.~\cite{KP04} is
also shown as a shaded area.
}
\label{fig:f2d-final}
\end{center}
\end{figure}

\subsection{Systematic Studies}
\label{sec:sys}

As we discuss in Sec.~\ref{sec:deltaf}, the uncertainty on the off-shell function $\delta f(x)$ 
related to the modeling of the deuteron wave function turns out to be negligible compared to the 
statistical uncertainty from our global QCD fits. 
Our final $\pm 1 \sigma$ band on the ratio $F_2^D/F_2^N$, computed 
by averaging the results from the five individual fits with 
different wave functions in Fig.~\ref{fig:f2d}, is given in Fig.~\ref{fig:f2d-final}, including 
the statistical uncertainty from the fit and the one related to the choice of the
deuteron wave function added in quadrature. 

In Ref.~\cite{Alekhin:2007fh} it was shown that there is some tension between the DIS data sets from the 
BCDMS and SLAC experiments, resulting in significant modifications of the extracted HT terms and PDFs. 
In order to mitigate the impact of this tension on our studies we allow the overall normalization of both 
the BCDMS proton and deuteron data sets to vary freely in our fits. This approach is justified by the use of separate 
normalizations for the deuterium and proton data sets in the BCDMS measurements~\cite{Benvenuti:1989rh,Benvenuti:1989fm}. 
The normalization of the BCDMS proton data is essentially defined by the precise HERA data in the overlap 
region, resulting in an overall factor consistent with the
corresponding normalization uncertainties quoted by the experiment, up to 3\%.  
The only constraint on the normalization of the BCDMS deuteron data comes from the SLAC 
experiments. However, the partial correlation between this normalization and the determination of the  
deuteron nuclear correction can potentially introduce an additional uncertainty in the global fits. 
The recent direct measurement of the deuteron nuclear correction by the 
BONuS experiment substantially reduces this uncertainty by constraining the normalization of the 
overall nuclear corrections. As a result, the normalization factor for the BCDMS deuterium data 
obtained from our fits is stable against variations of the deuteron wave function and is very 
close to the one obtained for the BCDMS proton data.    

As an additional test of the robustness of our analysis, we perform separate fits to different subsets 
of the data listed in Table~\ref{tab:data}. These variants of our analysis do not indicate any 
anomalous tension related to individual data samples, but rather suggest that our results on the 
off-shell correction are originated from the combined fit of all deuteron data sets.

\begin{figure}[tb] 
\begin{center}
\includegraphics[width=1.0\textwidth]{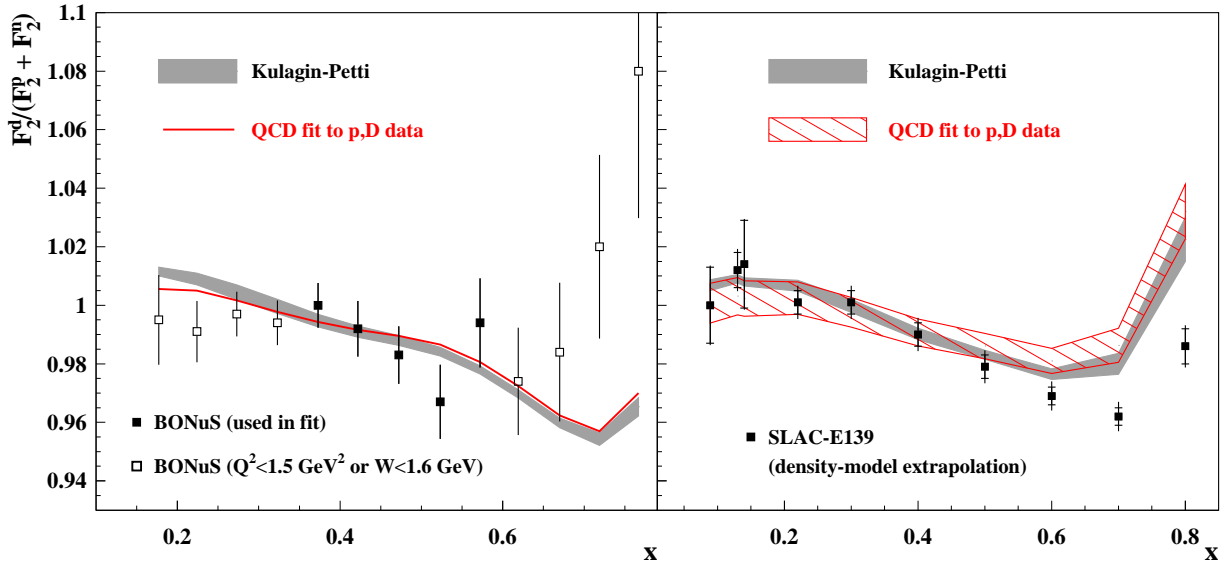}
\caption{%
Left panel: The ratio $F_2^D/F_2^N$ measured by the BONuS experiment~\cite{Griffioen:2015hxa} compared to  
the central value obtained from this analysis (solid line), and to the $\pm 1 \sigma$ uncertainty associated 
with the predictions from Ref.~\cite{KP04} (shaded area).  
Right panel: The model-dependent extrapolation of 
heavy target data from SLAC E139~\cite{Gomez:1993ri} within the nuclear density model~\cite{Frankfurt:1988nt} 
compared to the corresponding $\pm 1 \sigma$ uncertainty band from this analysis (hatched band) and to 
the predictions from Ref.~\cite{KP04} (shaded area). 
}
\label{fig:f2d-bonus}
\end{center}
\end{figure}

\subsection{Discussion}
\label{sec:discussion}

The results of our analysis discussed in Sec.~\ref{sec:deltaf} support the predictions 
of Ref.~\cite{KP04} for the nuclear effects in the deuteron and the unified treatment of all nuclei.  
We can then exploit the higher precision offered by the existing DIS data off heavier nuclear 
targets ($A\geq4$) to fix the value of the off-shell function $\delta f$ used in 
global QCD fits following Ref.~\cite{KP04}. 
The corresponding reduction of the overall uncertainties 
on the deuteron nuclear corrections is illustrated in Fig.~\ref{fig:f2d-final}. 
Within a simple single-scale model, relating the quark momentum distributions in the 
nucleon to the nucleon radius~\cite{KP04}, the off-shell function $\delta f$ 
obtained from nuclei with $A\geq 4$ 
suggests an increase of the nucleon core radius by about 2\% in the deuteron, taking  
an average virtuality of $-0.045$ from Table~\ref{tab:kinwf}. This value is comparable 
to estimates obtained with a different model~\cite{Close:1984zn} in relation to an 
increase of the overlap of nucleons in nuclei with the nuclear density.  

The predictions from both the present analysis and the ones from Ref.~\cite{KP04} are compared 
with the recent direct measurement of the
ratio $F_2^D/F_2^N$ by the BONuS experiment~\cite{Griffioen:2015hxa} in Fig.~\ref{fig:f2d-bonus}.  
A good agreement with BONuS data is observed, although the current experimental accuracy is 
somewhat limited and many data points fall in the resonance region with $W<1.6$ GeV or 
correspond to low $Q^2<1.5$ GeV$^2$.
The model-dependent extrapolation of the $F_2^A$ measurements with $A\geq4$ performed by the SLAC E139
experiment~\cite{Gomez:1993ri} is also compared with our results and with the ones from Ref.~\cite{KP04} 
in Fig.~\ref{fig:f2d-bonus}.  
Although the basic assumption of Ref.~\cite{Gomez:1993ri} about the scaling of the magnitude of 
nuclear effects with the nuclear density~\cite{Frankfurt:1988nt} was excluded by the recent 
measurements on ${}^9$Be target by the E03-103 experiment~\cite{Seely:2009gt}, the 
results of Ref.~\cite{Gomez:1993ri} with $x<0.7$ are consistent with our predictions.    
It is worth noting that the typical $Q^2$ values of the E139 data are  
substantially larger than in the BONuS sample (2-15 GeV$^2$ vs. 1-4 GeV$^2$). 
This difference allows to demonstrate the $Q^2$ dependence in the nuclear corrections, 
which appears mainly at large $x$ due to the combined effect of the  
TMC and the off-shell corrections, cf. Fig.~\ref{fig:f2d-bonus}. 

Since nuclear corrections are almost linearly dependent from $x$ in the 
region $0.35 < x < 0.55$, they are often quantified by the corresponding linear slope. 
The main advantage of this slope is that it can be measured more accurately since it 
is not affected by the normalization uncertainties. 
The empirical model-independent determination of the slope 
$\ud {\mathcal R}(D/N) / \ud x=-0.100\pm0.050$~\cite{Griffioen:2015hxa} of the BONuS data  
agrees well with the value $-0.099\pm0.006$ predicted by Ref.~\cite{KP04}.
To this end, the model-dependent extrapolation of the SLAC E139 data~\cite{Gomez:1993ri}  
gives a consistent value of $-0.098\pm0.005$, while 
the empirical extrapolation using the short range correlation scale factors 
from Ref.~\cite{Weinstein:2010rt} results in a somewhat smaller slope $-0.079\pm0.006$.
However, while being useful for the analysis of experimental data, the slope $\ud {\mathcal R}(D/N) / \ud x$  
describes the behavior of the nuclear corrections in a limited region only. 
Meanwhile, the microscopic model of Ref.~\cite{KP04} reproduces not only the measured slopes,  
but also the shape and magnitude of the nuclear corrections in the entire kinematic range 
covered by existing data. 

Our results on the off-shell correction differ from the ones obtained 
using a similar formalism in Refs.~\cite{Accardi:2011fa,Accardi:2016qay}.  
The analysis of Ref.~\cite{Accardi:2011fa} is based on a modified model of Ref.~\cite{KP04} 
to relate the off-shell function $\delta f$ to an increase in the nucleon confinement radius 
in the nuclear medium. The analysis of Ref.~\cite{Accardi:2016qay} follows more closely the 
model of Ref.~\cite{KP04}, with the off-shell function $\delta f$ being detrmined from a global 
QCD fit to the deuteron and proton data. The differences in the results of those analyses 
with respect to our study can be attributed to the implementation of the deuteron model and 
the details of the calculations~\cite{AccardiDIS16}, as discussed in the Appendix. 

As discussed in Sec.~\ref{sec:deltaf}, the ratio of DIS structure functions for two different 
nuclear targets, $\mathcal R(A/B) = F_2^A/F_2^B$, offers a good tool to  
study the off-shell function $\delta f(x)$, due to a large  
cancellation of both experimental and model uncertainties. In this respect we 
note that the addition of data on nuclear ratios $\mathcal R(A/B)$ to global QCD fits  
should improve the determination of the nucleon off-shell function $\delta f$ 
and of the $d$ quark distribution in the proton.

\begin{figure}[htb] 
\begin{center}
\includegraphics[width=0.8\textwidth]{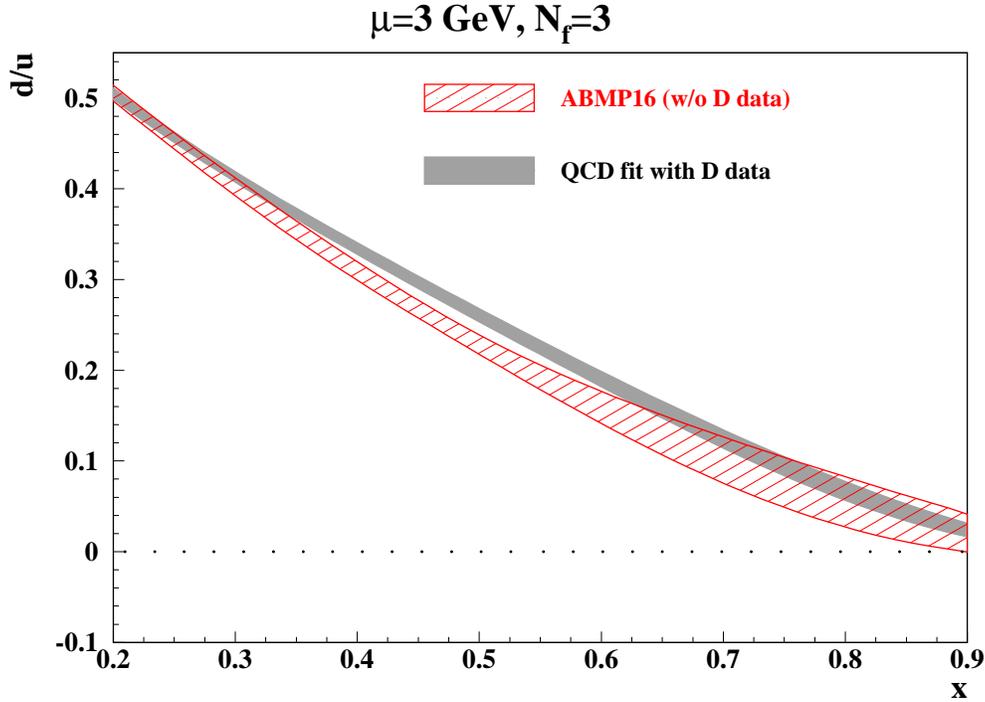}
\caption{%
Ratio $d/u$ at a factorization scale $\mu=3$ GeV as a function of $x$ 
obtained from global QCD fits. The hatched $\pm 1\sigma$ 
error band corresponds to the ABMP16 fit~\cite{Alekhin:2017kpj}, which does not include deuteron data.  
The shaded band shows the corresponding results obtained by including DIS deuteron data and by using the 
off-shell function $\delta f(x)$ and its uncertainty determined in Ref.~\cite{KP04}.  
}
\label{fig:d_u}
\end{center}
\end{figure}

\subsection{Constraints on $d/u$ and $F_2^n/F_2^p$}
\label{sec:duratio}

Correlations between the deuteron nuclear corrections and the $d$-quark distribution  
can substantially limit the PDF accuracy achievable in the PDF fits based on the 
proton and deuterium DIS data. 
In this context the data from flavor sensitive processes like $W^\pm$ production in 
$pp(\bar{p})$ collisions play a major role in reducing such correlations. 
A possible approach to avoid the effects of the deuteron nuclear corrections is to avoid any 
DIS data off the deuteron in global QCD fits, as in the recent ABMP16 analysis~\cite{Alekhin:2017kpj}. 
The corresponding results for the $d/u$ ratio shown in Fig.~\ref{fig:d_u} 
indicate that the recent precision data on $W^\pm$ boson production
from D0 and the LHC experiments (cf. Table~\ref{tab:data}) provide a good sensitivity 
to the $d$-quark distribution. 
In particular, the $d/u$ ratio at large $x>0.7$ is well constrained, mainly due to the large rapidity data 
from the recent LHCb measurement of $W^\pm$ boson production~\cite{Aaij:2015gna,Aaij:2015zlq}.  
This sample indeed probes values of $x$ up to 0.8 and its accuracy is comparable to the one of DIS experiments. 

The universality of $\delta f$ allows a further improvement of the accuracy in the determination of the 
$d/u$ ratio, by using the deuteron DIS data in combination with the more precise
off-shell function obtained from the analysis of the nuclear targets with $A\geq4$ (see Sec.~\ref{sec:deltaf}).
In Fig.~\ref{fig:d_u} we show the $d/u$ ratio obtained in such a way in the ABMP16 fit. 
The impact of the DIS deuteron data on the $d/u$ ratio is more evident in the region of $x>0.4$, 
where the uncertainties are substantially reduced, 
as compared to the ABMP16 results obtained without the DIS deuteron data.  

An interesting observation is that the $d/u$ ratio tends to vanish as $x \to 1$ (see Fig.~\ref{fig:d_u}). 
In order to verify that this behavior is not biased by the functional form of the 
PDF parametrization, we multiply the $d$- and $u$-quark parametrizations by an additional free polynomial. 
We do not find any significant impact on the corresponding large-$x$ behavior 
of the $d/u$ ratio, thus confirming its stability. 
Furthermore, the value of $d/u$ obtained in the present analysis is
consistent with the one of the ABMP16 fit performed without using the
deuteron data, cf. Fig.~\ref{fig:d_u}. These results indicate that our $d$- and $u$-quark 
parametrizations are flexible enough to be driven by the data sets, 
rather than by the functional form used (cf. Appendix).

\begin{figure}[tb] 
\begin{center}
\includegraphics[width=0.8\textwidth]{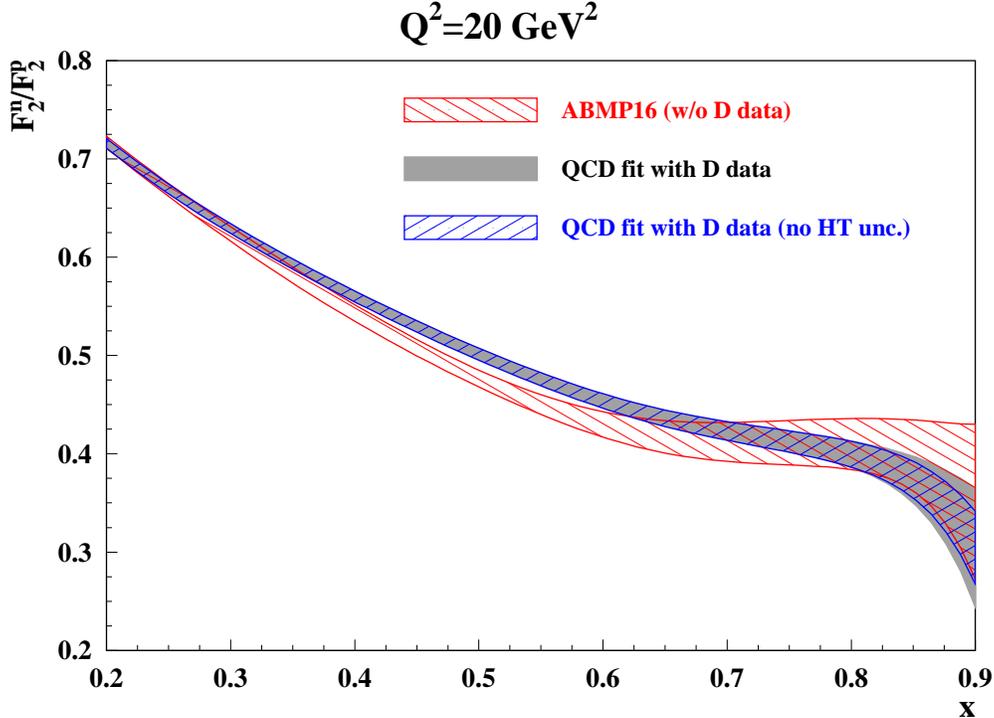}
\caption{%
The same as in Fig.~\ref{fig:d_u}, for the ratio $F_2^n/F_2^p$ as a function of $x$ at $Q^2=20\gevsq$. 
The $\pm 1 \sigma$ error band corresponding to the QCD fit with the deuteron data included and no 
HT uncertainties taken into account is also given for comparison (right-tilted hatch). 
}
\label{fig:n_p}
\end{center}
\end{figure}

The $d/u$ ratio is related to the neutron to proton structure function ratio,     
$F_2^n/F_2^p$, displayed in Fig.~\ref{fig:n_p}. 
The impact of the deuteron nuclear correction on $F_2^n/F_2^p$ is somewhat larger 
than on the $d/u$ ratio. Note, however, that the behavior of $F_2^n/F_2^p$ at $x \to 1$ is 
dominated by the HT contributions, which introduce a significant uncertainty on this ratio, 
as demonstrated in Fig.~\ref{fig:n_p}.

\begin{figure}[tb] 
\begin{center}
\includegraphics[width=0.8\textwidth]{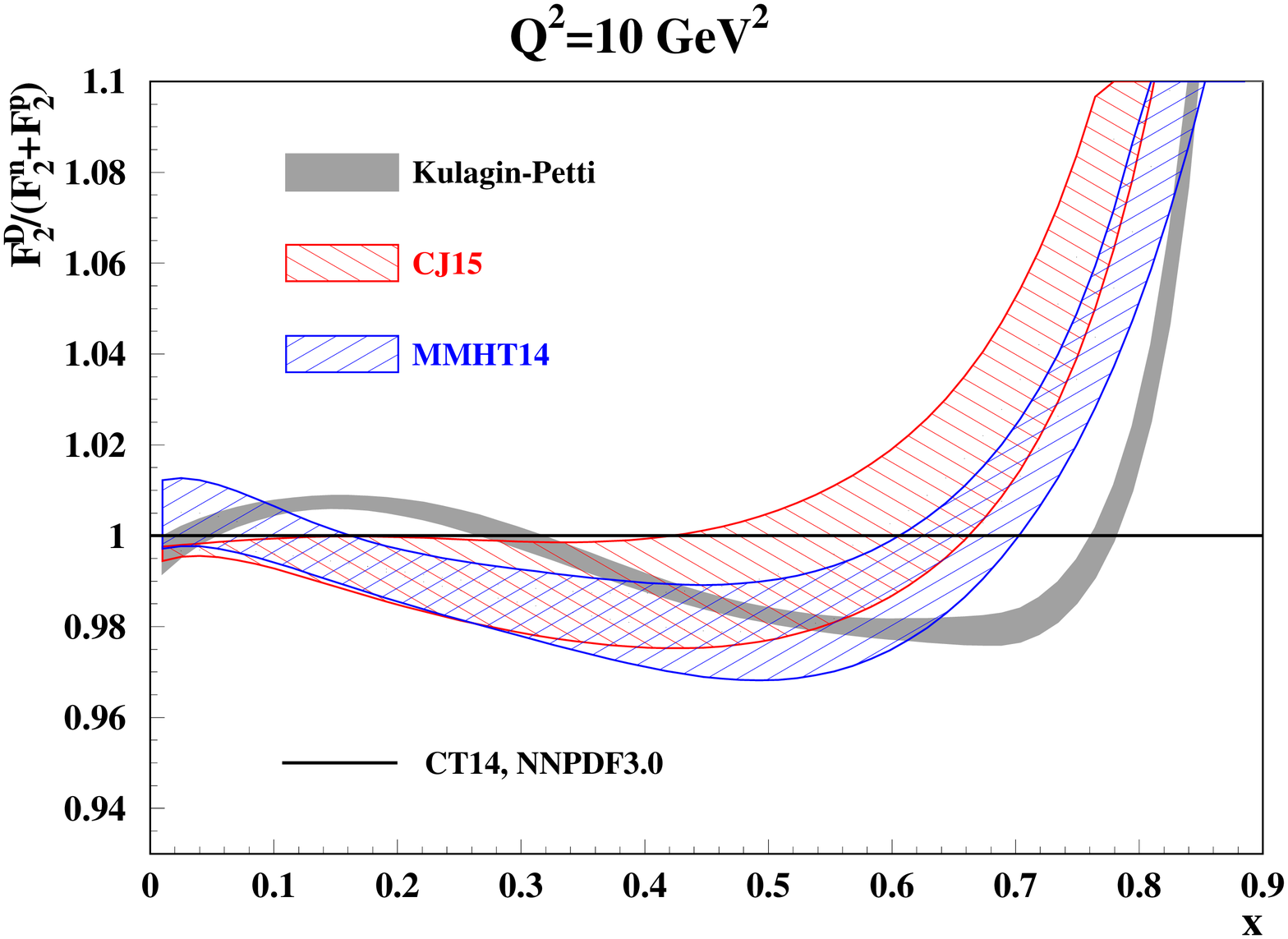}
\caption{%
Comparison of the $F_2^D/F_2^N$ parametrization with the corresponding uncertainties used to correct 
for nuclear effects in the  
CJ15~\cite{Accardi:2016qay}, MMHT14~\cite{Harland-Lang:2014zoa}, CT14~\cite{Dulat:2015mca}, 
and NNPDF3.0~\cite{Ball:2014uwa} analyses.  
The solid gray band gives the $\pm 1 \sigma$ uncertainty associated with the predictions of Ref.~\cite{KP04}.
}
\label{fig:f2d-PDFsFits}
\end{center}
\end{figure}

It is instructive to compare the nuclear corrections applied to the deuteron data in various PDF analyses. 
The CJ15 analysis~\cite{Accardi:2016qay} is based upon a formalism similar to 
the model of Ref.~\cite{KP04}, with the corresponding off-shell correction 
in the deuteron determined from the global QCD fit (see discussion in Sec.~\ref{sec:discussion}).  
The MMHT14 analysis~\cite{Harland-Lang:2014zoa} is based on an empirical parametrization 
of the nuclear correction to the ratio $F_2^D/F_2^N$, which is extracted from data. 
The CT14~\cite{Dulat:2015mca} and NNPDF3.0~\cite{Ball:2014uwa} do not use any nuclear 
correction arguing that nuclear corrections would introduce additional  
uncertainties in the analysis~\cite{Ball:2014uwa}. Furthermore, it was claimed that 
nuclear corrections can be neglected 
when using more stringent cuts in $Q^2$ and $W^2$~\cite{Dulat:2015mca}. 
The nuclear corrections used by CJ15 and
MMHT14, while mutually consistent, are characterized by relatively large uncertainties 
(cf. Fig.~\ref{fig:f2d-PDFsFits}). 
Since these two determinations are driven by the available deuterium data and are 
largely correlated with the $d/u$ ratio, the possibility to improve such uncertainties 
in the context of global QCD fits appears to be limited. 
As discussed above, the use of the microscopic model of Ref.~\cite{KP04} 
and of the corresponding results on the universal off-shell function $\delta f$, allows 
a substantial reduction of the uncertainties related to the nuclear correction in the deuteron. 
Such a reduction is illustrated by the comparison with the model
of Ref.~\cite{KP04} given in Fig.~\ref{fig:f2d-PDFsFits}. 
In turn, the uncertainty on the deuteron nuclear correction is a bottleneck in the 
overall accuracy of the $d/u$ ratio obtained from global QCD fits.  
The differences in the treatment of the nuclear corrections in the deuteron 
illustrated in Fig.~\ref{fig:f2d-PDFsFits} would translate into a corresponding spread   
in the $d$-quark distribution. 
It is worth noting that these systematic effects cannot be mitigated by more stringent 
$Q^2$ and $W^2$ cuts, since nuclear effects survive even 
at very large energy and momentum, as demonstrated in DIS experiments 
(for a review see Refs.~\cite{Arneodo:1992wf,Norton:2003cb})  
and by recent observations of nuclear modifications in 
$p+{\rm Pb}$ and ${\rm Pb}+{\rm Pb}$ collisions at the 
LHC~\cite{Khachatryan:2015hha,Khachatryan:2015pzs,Aad:2015gta,Aaij:2014pvu,Senosi:2015omk,Aad:2012ew,Chatrchyan:2014csa}.  

The largest deviations between the CJ15 and MMHT14 analyses and the model of Ref.~\cite{KP04} are 
observed in the intermediate region $x\sim 0.15$ and at large $x>0.6$ (cf. Fig.~\ref{fig:f2d-PDFsFits}), 
although their significance is limited by the current uncertainties.  
Our analysis of deuteron data is consistent with Ref.~\cite{KP04}, with an uncertainty band extending 
close to the CJ15 and MMHT14 ones (cf. Fig.~\ref{fig:f2d-final}). 
The small enhancement present in the model of Ref.~\cite{KP04} for $0.05<x<0.3$ is the result of 
an interplay of the off-shell correction and the meson exchange currents (cf. Fig.~\ref{fig:d_n}).  
Nuclear corrections at $x>0.6$ are instead dominated by the FMB and OS corrections.  
This kinematic region is very sensitive to the treatment of both the bound nucleon momentum distribution and 
the target mass corrections to the nucleon structure functions (Sec.~\ref{sec:model}).
It is worth noting that the prescription of Ref.~\cite{Georgi:1976ve} for the TMC is known to have an incorrect 
behavior for $x\to 1$, which can affect the calculations at very large $x$ values. 
More detailed comparisons with the CJ15 and MMHT14 results can be found in the Appendix. 
Future DIS measurements from the BONuS experiment with the 12 GeV JLab upgrade~\cite{BONuS12}, 
from neutrino and antineutrino scattering off free proton in the DUNE experiment~\cite{Acciarri:2015uup}, 
and from the electron-ion collider~\cite{Accardi:2012qut} can further 
improve our understanding of the ratio $F_2^D/F_2^N$.

\section{Summary} 
\label{sec:sum}

We performed a study of nuclear effects in the deuteron 
using the data from DIS off proton and deuterium, Drell-Yan production in 
$pp$ and $p{\rm D}$ interactions, and $W^\pm$- and $Z$-boson production in $pp$ and $p\bar{p}$ 
collisions in the context of global QCD fits. We found that it is possible to 
determine simultaneously PDFs, high twist terms, and the off-shell function describing the modification of 
PDFs in bound nucleons due to their different $Q^2$ dependence and the wide kinematic coverage 
of existing data. Flavor sensitive processes like $W^\pm$ production in $pp(\bar{p})$ collisions 
play an important role in disentangling the impact of the nuclear corrections in the deuteron 
from the $d$ quark distribution function, allowing a more accurate determination of both contributions. 
We also evaluated the sensitivity of our results to various models of the deuteron  
wave function and found that the corresponding model dependence is reduced by the 
recent BONuS measurement of the ratio $F_2^D/F_2^N$. 

The results on the off-shell function $\delta f$ reported in this paper are in good agreement 
with the earlier determination obtained in the analysis of the ratio of nuclear structure functions 
with mass number $A\geq4$~\cite{KP04}. This result confirms the universality of $\delta f$, 
which can be regarded as a special structure function 
of the nucleon describing the modification of the bound nucleons in the nuclear medium. 
This study supports the unified treatment of the deuteron and heavier nuclei 
developed in Ref.~\cite{KP04}. 

We also studied the impact of deuteron nuclear corrections on the $d/u$ ratio within global QCD fits. 
We found that the recent precision data on $W^\pm$ boson production from D0 and the LHC experiments 
allow a reduction of the uncertainties in the $d/u$ ratio at large $x$.
Our results indicate that the accuracy in the determination of the $d/u$ ratio can be further 
substantially improved by including the DIS data off a deuterium target corrected for nuclear effects 
using the model of Ref.~\cite{KP04} with the universal off-shell function $\delta f$.

\begin{acknowledgments}

We thank F. Gross and W. Polyzou for providing the parametrizations of the WJC and AV18 deuteron 
wave functions, respectively. 
We thank R. Machleidt for the tables of the CD-Bonn deuteron wave function. 
We thank A. Accardi, W. Melnitchouk, S. Moch, and J. Owens for fruitful discussions.   
R.P. thanks the II Institute for Theoretical Physics at the University of Hamburg for 
hospitality during the manuscript preparation. 
S.A. was supported by Bundesministerium f\"ur Bildung und Forschung (Contract No. 05H15GUCC1). 
R.P. was supported by Grant No. DE-SC0010073 from the Department of Energy, USA.
S.K. was supported by the Russian Science Foundation Grant No. 14-22-00161.

\end{acknowledgments}

\newpage 
\appendix* 
\section{Results from different analyses}

\begin{figure}[htb] 
\begin{center}
\includegraphics[width=0.75\textwidth]{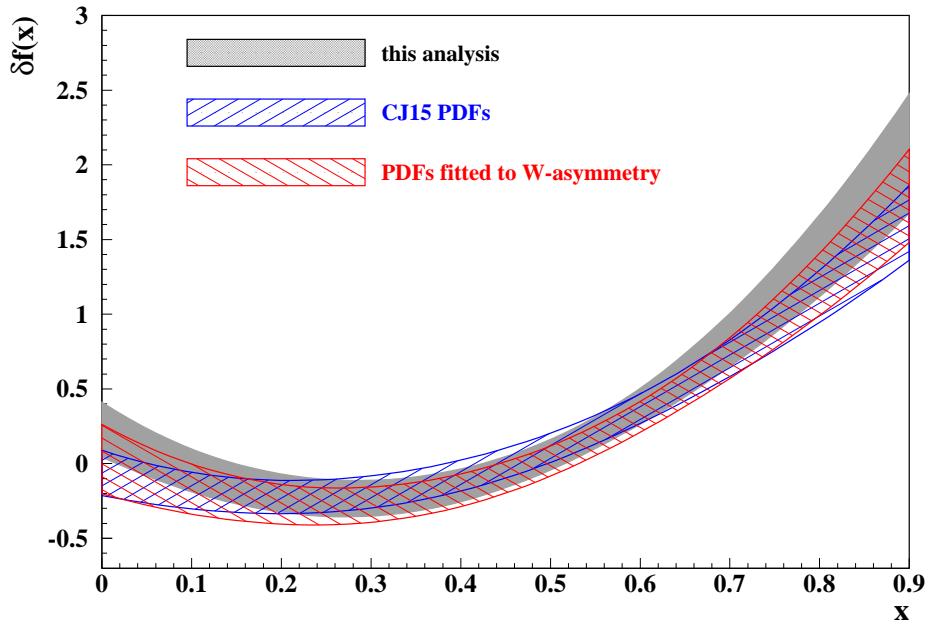}
\caption{%
Comparison of the off-shell functions $\delta f(x)$ ($\pm 1 \sigma$ uncertainty bands)
extracted from different global QCD fits including deuteron data  
(shaded area: our results with the AV18 deuteron wave function~\cite{Veerasamy:2011ak}; 
left-tilted hatch: result obtained by using the D0 data on W-boson asymmetry~\cite{Abazov:2013dsa}). 
The result from a NLO fit using fixed CJ15 proton PDFs from 
the LHAPDF library~\cite{Buckley:2014ana} is also displayed (right-tilted hatch). 
}
\label{fig:CJ15deltaf}
\end{center}
\end{figure}

A few phenomenological studies aimed to extract the nuclear corrections in the deuteron from 
global QCD fits are available in literature~\cite{Accardi:2011fa,Harland-Lang:2014zoa,Accardi:2016qay}, 
including the analyses of Refs.~\cite{Accardi:2011fa,Accardi:2016qay} based on a deuteron  
model similar to that used in our studies. 
In this Appendix we investigate the differences with respect to those studies~\cite{AccardiDIS16} and 
benchmark the recent CJ15 analysis~\cite{Accardi:2016qay}.   

Clear differences with the CJ15 fit appear in the data sets used and in the proton PDFs correspondingly 
obtained. From Table I of Ref.~\cite{Accardi:2016qay} we conclude that most 
of the sensitivity to the nuclear corrections in the CJ15 fit manifests for the D0 data on the $W^\pm$-boson 
production asymmetry and the DIS deuteron data by the SLAC experiments 
(cf. reduction of $\chi^2$ obtained by adding the nuclear corrections).
While we include the SLAC DIS data (Table~\ref{tab:data}), 
we use significantly different $W,Z$-boson collider data in our analysis  
(see Sec.~\ref{sec:data}). More specifically, we include the D0 data on the lepton asymmetry 
from the $W$-boson decays rather than the actual $W$-boson asymmetry data, as well as the recent LHC 
DY data~\cite{Aad:2011dm,Aad:2016naf,Chatrchyan:2013mza,Khachatryan:2016pev,Aaij:2015gna,Aaij:2015vua,Aaij:2015zlq}.  
In order to test the impact of such differences, we perform a variant of our fit in which we drop all the 
$W,Z$-boson collider data, replacing them with the D0 $W$-asymmetry data~\cite{Abazov:2013dsa}. 
The corresponding results for the function $\delta f$ are consistent with the ones 
presented in Sec.~\ref{sec:deltaf} (cf. Fig.~\ref{fig:CJ15deltaf}). 
We also perform a separate NLO fit to all deuteron data sets using fixed CJ15 proton PDFs from 
the LHAPDF library~\cite{Buckley:2014ana}. The off-shell function $\delta f$ obtained in this way  
is consistent with our results in Sec.~\ref{sec:deltaf} and Ref.~\cite{KP04} (cf. Fig.~\ref{fig:CJ15deltaf}).  
Therefore, we can conclude that the differences in the data samples and 
proton PDFs cannot explain the different off-shell correction obtained in 
Ref.~\cite{Accardi:2016qay}. 

\begin{figure}[tb] 
\begin{center}
\includegraphics[width=0.5\textwidth]{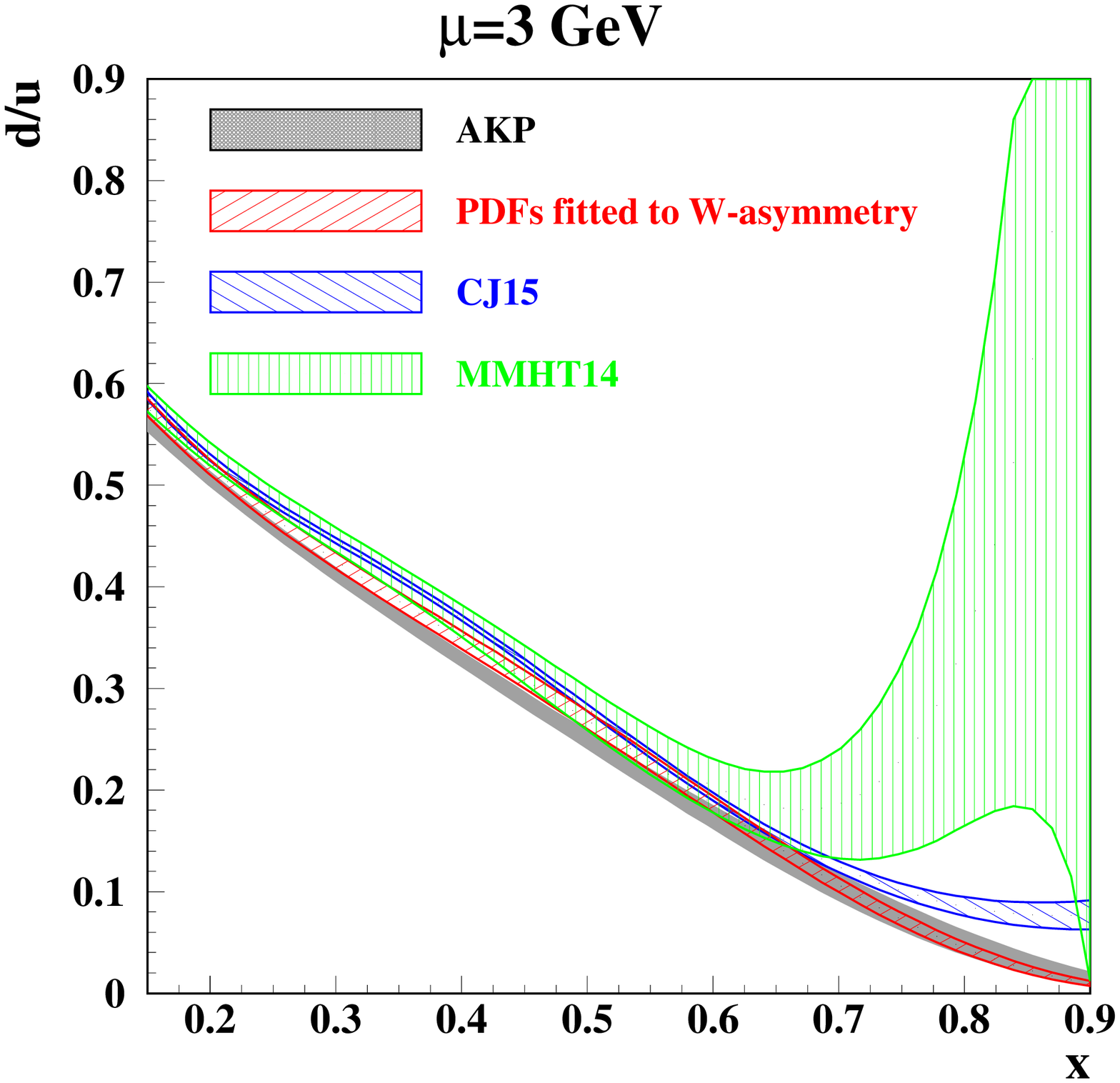}\includegraphics[width=0.5\textwidth]{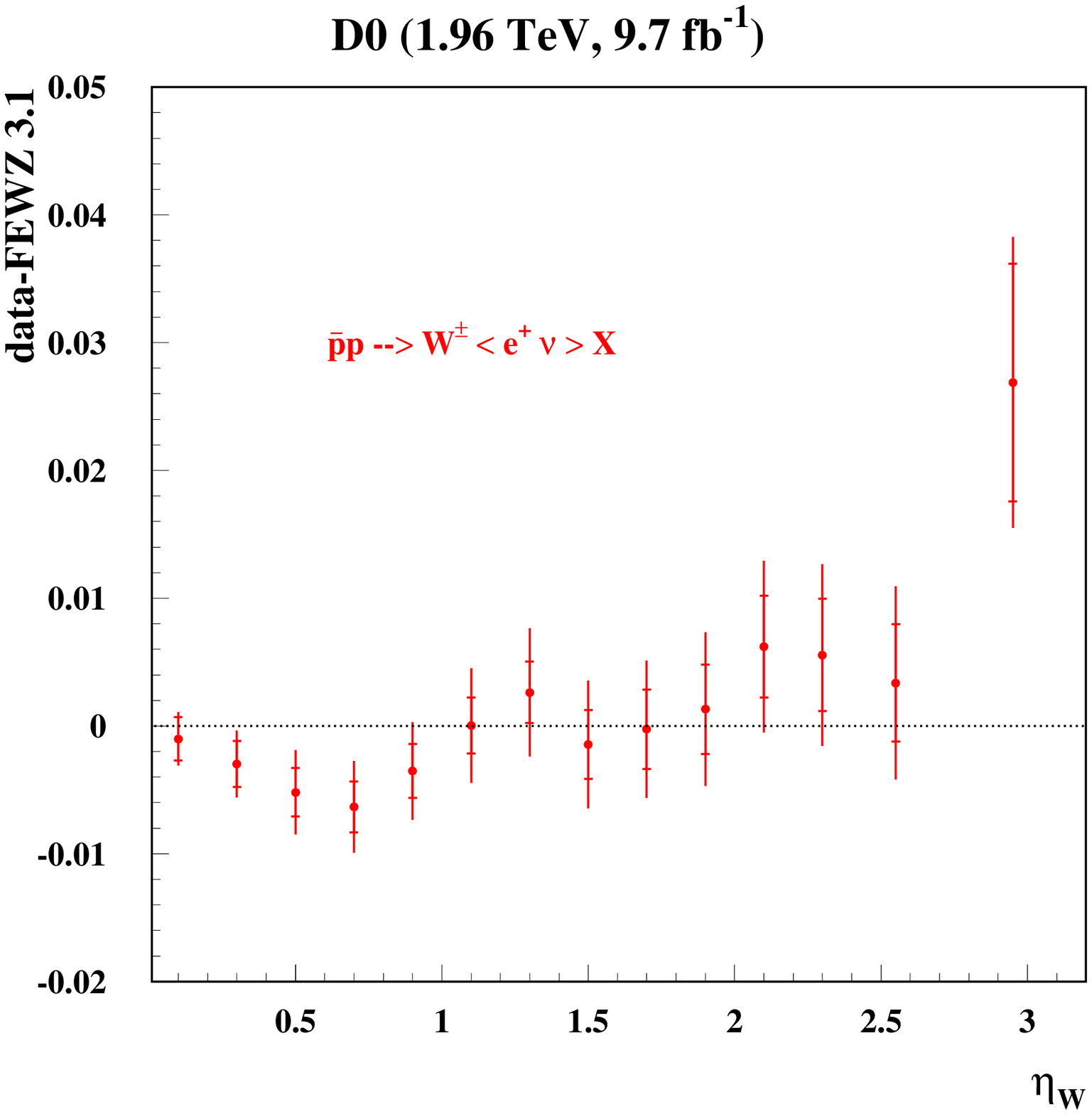}
\caption{%
Left panel: Comparison of the $d/u$ ratio ($\pm 1 \sigma$ uncertainty bands)
obtained in different global QCD fits 
(shaded area: present analysis; right-tilted hatch: variant of the our analysis using 
the D0 data on $W$-boson asymmetry~\cite{Abazov:2013dsa}; left-tilted hatch: 
NLO CJ15 fit~\cite{Accardi:2016qay}; vertical hatch: NNLO MMHT14 fit~\cite{Harland-Lang:2014zoa}). 
Right panel: Difference between the D0 W-asymmetry data~\cite{Abazov:2013dsa} and the NLO predictions obtained 
with the FEWZ package~\cite{Li:2012wna,Gavin:2012sy} using CJ15 
PDFs from the LHAPDF library~\cite{Buckley:2014ana}.  
}
\label{fig:CJ15du}
\end{center}
\end{figure}

Meanwhile, it is instructive to compare the values of the $d/u$ ratio obtained in the various fits.  
The $d/u$ ratio obtained from the variant of our fit with the D0 $W$-asymmetry data is 
consistent within uncertainties with the results presented in Sec.~\ref{sec:duratio} (cf. Fig.~\ref{fig:CJ15du}). 
The differences between the central values suggest that our PDF parametrizations are flexible enough 
to describe different data sets without limitations from the functional form used. 
In particular, we do not 
explicitly constrain the $d/u$ ratio to vanish for $x\to 1$.   
The flexibility of our PDF parametrization is confirmed by the fact that we obtain 
similar results by multiplying the $d$- and $u$-quark parametrizations by an 
additional free polynomial (cf. Sec.~\ref{sec:duratio}). 
The uncertainty of the $d/u$ ratio determined in the MMHT14 analysis~\cite{Harland-Lang:2014zoa} 
at large $x$ does not allow quantitative comparisons (cf. Fig.~\ref{fig:CJ15du}), 
due to the lack of relevant experimental constraints.
It is worth noting that the $d/u$ ratio of the CJ15 analysis~\cite{Accardi:2016qay} 
displays a substantially different behavior at large $x$. 
Since this kinematics is largely controlled by the data on $W$-boson asymmetry at large rapidity, 
we compute this quantity at the NLO approximation using the FEWZ 
package~\cite{Li:2012wna,Gavin:2012sy} with CJ15 proton PDFs from the LHAPDF library~\cite{Buckley:2014ana}. 
A comparison with the D0 data~\cite{Abazov:2013dsa} (c.f. Fig.~\ref{fig:CJ15du})  
shows deviations from these predictions at large values of the 
$W$-boson rapidity, in contrast with the corresponding results of the CJ15 analysis 
(cf. Fig.~13 of Ref.~\cite{Accardi:2016qay}).

\begin{figure}[tb] 
\begin{center}
\includegraphics[width=0.5\textwidth]{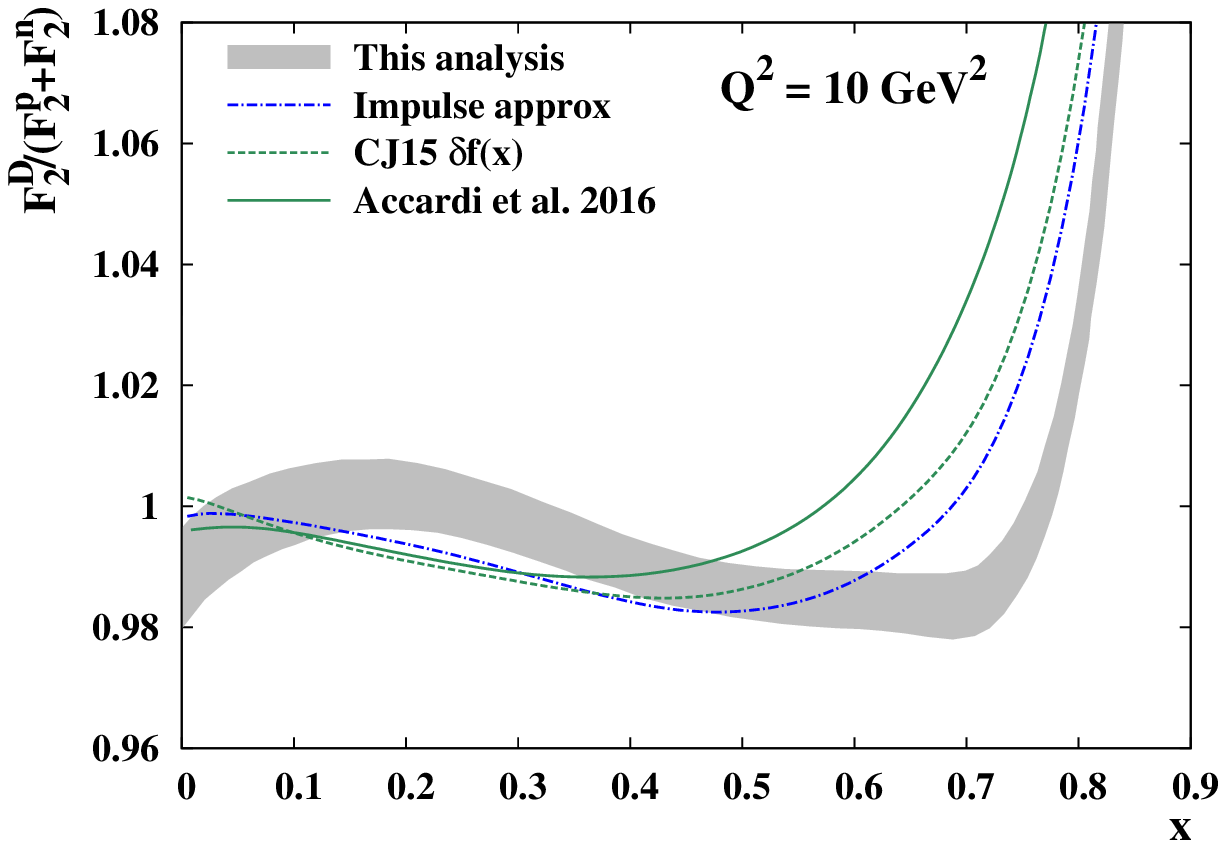}\includegraphics[width=0.5\textwidth]{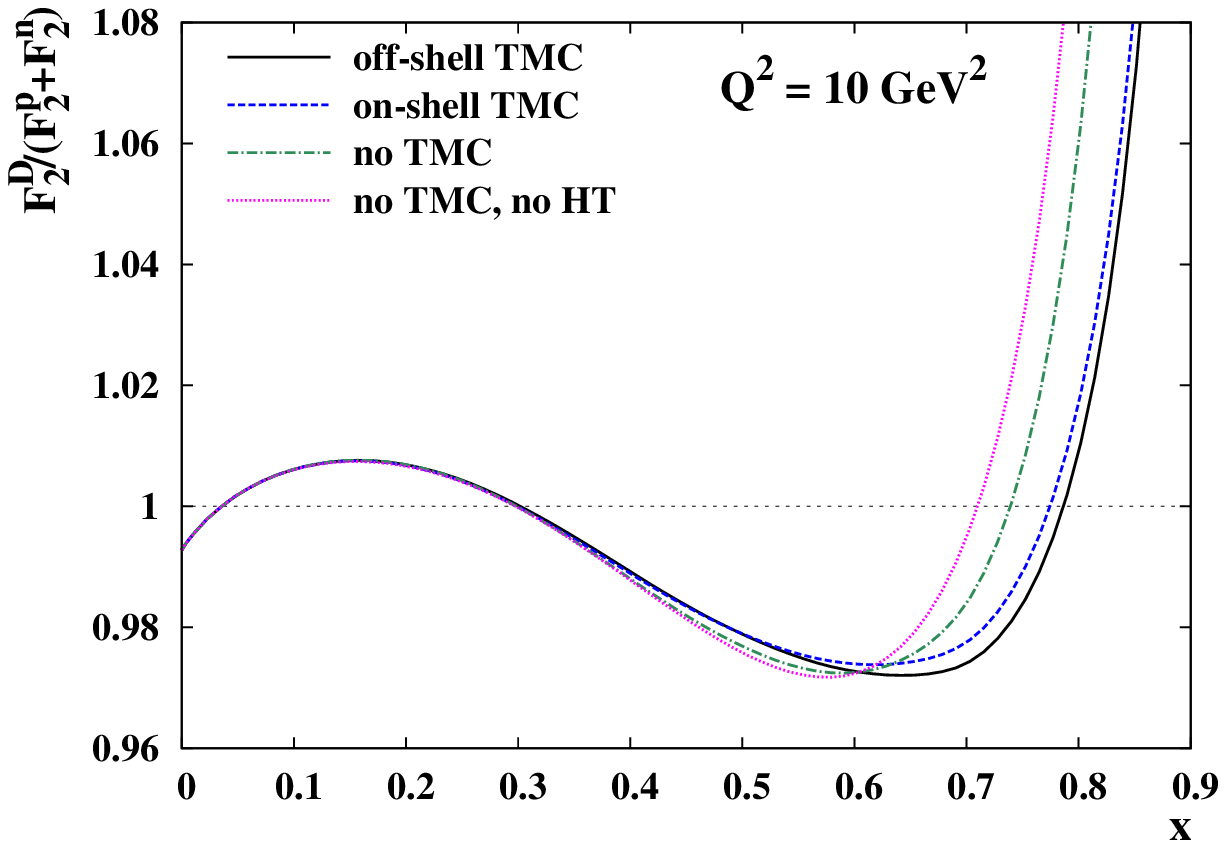}
\caption{%
Left panel: Comparison of the ratio $F_2^D/F_2^N$ from Ref.~\cite{Accardi:2016qay} (solid line) with 
our calculation using the same $\delta f(x)$ and AV18 deuteron wave function~\cite{Veerasamy:2011ak} 
(dashed line) and with the corresponding impulse approximation without OS correction (dash-dotted line).  
The $\pm 1\sigma$ uncertainty band obtained from this analysis is also shown as a shaded area. 
Right panel: Same ratio computed with the off-shell correction from Ref.~\cite{KP04} using  
different approximations: off-shell TMC from \eq{eq:TMC} and HT (solid line), 
on-shell TMC~\cite{Georgi:1976ve} and HT (dashed line), 
no TMC (dash-dotted line), no TMC nor HT (dotted line).  
}
\label{fig:CJ15-OS}
\end{center}
\end{figure}

A possible source of differences between this analysis and Ref.\cite{Accardi:2016qay}
may stem from the implementation of the convolution model, as well as from the 
treatment of the TMC and/or the HT contributions.
In order to check the sensitivity to these effects we compute the ratio
$F_2^D/F_2^N$ using the $\delta f(x)$ function from Ref.\cite{Accardi:2016qay} and
the AV18 wave function~\cite{Veerasamy:2011ak}.
The result shown in Fig.~\ref{fig:CJ15-OS} (left panel) deviates significantly from 
both the one presented in Sec.\ref{sec:deltaf} and the one of Ref.~\cite{Accardi:2016qay} 
using the same deuteron wave function, 
suggesting a different implementation of the convolution model in Ref.\cite{Accardi:2016qay}. 
Figure~\ref{fig:CJ15-OS} (right panel) also illustrates the sensitivity of the ratio $F_2^D/F_2^N$ to 
various implementations of the TMC and HT corrections.~\footnote{Both TMC and HT corrections are 
characterized by a strong $Q^2$ dependence, increasing their impact at lower $Q^2$ values.} 
In particular, we compare the standard TMC scheme of Ref.~\cite{Georgi:1976ve} (on-shell TMC) 
with its off-shell continuation by \eq{eq:TMC}. 
The TMC and HT corrections to $F_2^D/F_2^N$ are treated differently in the CJ15 and MMHT14 analyses, 
and could contribute to the disagreement in the region of large $x$ (cf. left panel of Figure~\ref{fig:CJ15-OS}).

In summary, the present studies indicate that we cannot reproduce the CJ15 results of 
Ref.~\cite{Accardi:2016qay} on the function $\delta f$. 
All our systematic checks are consistent with the determination presented in Sec.~\ref{sec:deltaf}.  
The differences with the results of Ref.\cite{Accardi:2016qay} 
cannot be explained by the different data samples nor by the PDFs used. 
Instead, we find indications pointing towards the implementation of the deuteron model, 
TMC and HT corrections in the CJ15 fit.

\bibliography{main}

\end{document}